\documentclass[11pt,a4paper]{article}
\usepackage[utf8]{inputenc}

\usepackage{amsmath}
\usepackage{graphicx}
\usepackage{amssymb}
\usepackage{tabularx}
\usepackage{adjustbox}
\usepackage{csquotes}
\usepackage{comment}
\usepackage{imakeidx}
\usepackage{multirow}
\usepackage{xcolor}
\usepackage[margin=2.5cm]{geometry}
\definecolor{darkred}{rgb}{0.5, 0, 0}
\usepackage{hyperref}
\hypersetup{
     colorlinks=true,
     linkcolor=blue,
     filecolor=blue,
     citecolor=darkred,      
     urlcolor=cyan,
     }
\usepackage[toc, acronym, nopostdot, nonumberlist]{glossaries}
\usepackage{titling}
\usepackage{tikzpagenodes}
\usepackage[ddmmyyyy]{datetime}
\usepackage{setspace}
\usepackage{indentfirst}
\usepackage{lipsum}
\usepackage{cleveref}
\usepackage{makecell}
\usepackage{pdfpages}
\usepackage{geometry}
\usepackage{todonotes}
\usepackage[utf8]{inputenc}
\usepackage{booktabs}
\usepackage[numbers, sort&compress]{natbib}
\usepackage{authblk}
\usepackage{xurl}
\usepackage{setspace}

\makeglossaries
\title{\bfseries A typology of quantum algorithms}

\author[1,*]{Pablo Arnault}
\author[1]{Pablo Arrighi}
\author[2]{Steven Herbert}
\author[1,$\dag$]{Evi Kasnetsi}
\author[1]{Tianyi Li}

\affil[1]{\small Université Paris-Saclay, INRIA, CNRS, ENS Paris-Saclay, LMF, 91190 Gif-sur-Yvette, France}
\affil[2]{Quantinuum, Terrington House, 13-15 Hills Rd, Cambridge, CB2 1NL UK\vspace{-2.5mm}\\
Department of Computer Science and Technology, University of Cambridge, UK \vspace{0.4cm}} 

\affil[*]{pablo.arnault@universite-paris-saclay.fr}
\affil[$\dag$]{eviksnn@gmail.com}

\begin{document}

\maketitle

%\newgeometry{top=1cm, bottom=1cm, left=1cm, right=1cm}
%\input{acro_list}
%\restoregeometry

\pagenumbering{arabic}
%\pagenumbering{roman} 

%%%%%%%%%%%%%%%%%%%%%%%%%%%%%%%%%%%%%%%%%%%%%%%%%%%%%  
\begin{abstract}
%context

%problem

%proposal

%solution
\noindent
We draw the current landscape of quantum algorithms, by classifying about 130 quantum algorithms, according to the fundamental mathematical problems they solve, their real-world applications, the main subroutines they employ, and several other relevant criteria. The primary objectives include revealing trends of algorithms, identifying promising fields for implementations in the NISQ era, and identifying the key algorithmic primitives that power quantum advantage.

\vspace{2cm}

\end{abstract}

%\glsaddall
%\printglossary[type=\acronymtype]
%\printglossary[type=main,title={Glossary},toctitle={Glossary}]
%%%%%%%%%%%%%%%%%%%%%%%%%%%%%%%%%%%%%%%%%%%%%%%%%%%%%  

%\chapter{Introduction}
%\pagenumbering{arabic} 
%\label{chap:introduction}

%IMPORTANT:
%\begin{itemize}
%    \item PLEASE check the file vars.tex: I'm using it in 10\kbps.
%    \item Using glossaries for the first time: \gls{wn}
%    \item Using glossaries after first time: \gls{wn}
%    \item Using glossaries in plural: \glspl{wn}
%    \item Using glossaries extended: \glsdesc{wn}
%\end{itemize}

%How to reference and insert a \Cref{fig:labeloffigure}. 

%This is an example of a citation \cite{Pedro2019}.   

%This is a way for you to generate tables: \href{https://www.tablesgenerator.com/}%{https://www.tablesgenerator.com/}

%This is how to refer to another \Cref{cap:work}
%
%Here is how to do a bullet list
%\begin{itemize}
%    \item First point
%    \item Second point
%    \item and so on...
%\end{itemize}
%
%Here is how to do a numbered list
%\begin{enumerate}
%    \item First point
%    \item Second point
%    \item and so on...
%\end{enumerate}

%\section{Presentation of the host organization} 
%\label{sec:context}
%\noindent
%My internship was conducted in the Quantum Computation Structures (QuaCS) team, which is a %part of the Laboratoire Méthodes Formelles (LMF). The focus of team is .
\newpage
\vspace{-0.5cm}
\begin{center}
\textbf{\Large{Abbreviations (of single words)}} \\[7pt]
\end{center}
\vspace{-0.1cm}
\begin{tabular}{p{7.8cm}p{9cm}}
\parbox[c][0.8cm][c]{7.8cm}{\textbf{Q.} \ Quantum} \\[-2pt]
\parbox[c][0.8cm][c]{7.8cm}{\textbf{Algo.}  \ Algorithm}  \\[-2pt]
\end{tabular}
\\[0pt]

\vspace{-0.5cm}
\begin{center}
\textbf{\Large{Acronyms}}\\[7pt]
\end{center}
\vspace{-0.1cm}
\begin{tabular}{p{7.8cm}p{9cm}}
%\hline
\parbox[c][0.8cm][c]{7.8cm}{\textbf{Algorithms:}}& \parbox[c][0.8cm][c]{8.5cm}{\textbf{Families of algorithms:} } \\[-2pt]
\parbox[c][0.8cm][c]{7.8cm}{\textbf{QFT} \  Q.\   Fourier Transform} &\parbox[c][0.8cm][c]{8.5cm}{\textbf{HS}  \  Hamiltonian Simulation} \\[-2pt]
%\hline
\parbox[c][0.8cm][c]{7.8cm}{\textbf{AA}  \ Amplitude Amplification} &   \parbox[c][0.8cm][c]{8.5cm}{\textbf{QLSA}  \ Q.\ Linear-System Algo.\ }   \\[-2pt]
%\hline
\parbox[c][0.8cm][c]{8.5cm}{\textbf{QA}  \ Q.\ Adiabatic} &   \parbox[c][0.8cm][c]{8cm}{\textbf{VQA}  \  Variational Q.\ Algo.}  \\[-2pt]
 \parbox[c][0.8cm][c]{8.5cm}{\textbf{VQE}  \  Variational Q.\ Eigensolver}  &  \\[-2pt]
\parbox[c][0.8cm][c]{8.5cm}{\textbf{BS}  \  Boson Sampling} & \textbf{Others:}  \\[-2pt]

\parbox[c][0.8cm][c]{8.5cm}{\textbf{GBS}  \  Gaussian Boson Sampling} & \parbox[c][0.8cm][c]{8.5cm}{\textbf{NISQ}  \ Noisy Intermediate-Scale Quantum}  \\[-2pt]
\parbox[c][0.8cm][c]{7.8cm}{\textbf{QPE}  \ Q.\ Phase Estimation} & \parbox[c][0.8cm][c]{8.5cm}{\textbf{LSQ} \  Large-Scale Quantum} \\[-2pt]
 \parbox[c][0.8cm][c]{8.5cm}{\textbf{QAE}  \ Q.\ Amplitude Estimation}   &  \parbox[c][0.8cm][c]{8.5cm}{\textbf{QCA}  \ Q.\ Cellular Automaton}   \\[-2pt]
\parbox[c][0.8cm][c]{8.5cm}{\textbf{QSVT}  \ Q.\ Singular-Value Transformation} &  \parbox[c][0.8cm][c]{8.5cm}{\textbf{QW}  \ Q.\ Walk}   \\[-2pt]
\parbox[c][0.8cm][c]{8.5cm}{\textbf{QSP}  \ Q.\ Signal Processing} & \parbox[c][0.8cm][c]{9cm}{\textbf{HSP} \  Hidden-Subgroup Problem}  \\[-2pt]
%\hline
\parbox[c][0.8cm][c]{7.8cm}{\textbf{QAOA}  \ Q.\ Approximate Optimization Algo.} &   \\[-2pt]
%\hline

 \parbox[c][1.5cm][c]{8.5cm}{\textbf{QUBO}  \ Quadratic Unconstrained Binary \\ {\color{white} uuuuu \, \,} Optimization}  & \\[-2pt]
%\hline
 \parbox[c][0.8cm][c]{8.5cm}{\textbf{QITE}  \ Q.\ Imaginary-Time Evolution} & \\[-2pt]
\parbox[c][0.8cm][c]{8.5cm}{\textbf{LCU}  \ Linear Combination of Unitaries}  &   \\[-2pt]
%\hline
\parbox[c][0.8cm][c]{8.5cm}{\textbf{QSDP}  \ Q.\ Semi-Definite Programming} &   \\[-2pt]
%\hlineQAE
\parbox[c][0.8cm][c]{8.5cm}{\textbf{QSVD}  \ Q.\ Singular-Value Decomposition} &   \\[-2pt]
%\hlineQAE
\parbox[c][0.8cm][c]{8.5cm}{\textbf{HHL}  \ Harrow-Hassidim-Lloyd} &   \\[-2pt]
%\hlineQAE

%\hline
\end{tabular}
\\[5pt]
\vspace{0.8cm}

\begin{spacing}{1.0}
{\small
\noindent
{\bfseries Note 1:}   The above {\bfseries Abbreviations} will be used almost exclusively for names or descriptions of algorithms. The above {\bfseries Acronyms} will be used almost systematically, except when we mention the concept for the first time in the main text (not if this is done in the figures), or when we want to insist on, or recall the full name of the concept. Both the {\bfseries Abbreviations} and the {\bfseries Acronyms} are used both in the body of the paper and in the classification table of the appendix.} \end{spacing}

\vspace{1cm}
\begin{spacing}{1.0}
{\small
\noindent
{\bfseries Note 2:}  In this work, we choose to limit as much as possible the use of acronyms to names of algorithms, in order not to confuse the reader. As one can see above, there are however a few exceptions. But, in particular, we have not used acronyms for computational models (the acronyms QCA and QW above should hence not be understood as acronyms of computational models but rather of certain types or frameworks for certain algorithms), and similarly for the mathematical classes and the application domains that we define in our work.}
\end{spacing}

\newpage
\section{Introduction}
%\label{sec:motivation}
%\lipsum[1]
\noindent
\noindent
Quantum computing, an innovative and rapidly evolving field, holds the potential to reshape conventional paradigms in computation, by exploiting quantum-mechanical phenomena. One cornerstone of this transformation are quantum algorithms, whose design must enable a quantum computer to execute calculations more efficiently than classical computers. As the field thrives, the need for categorizing the multitude of quantum algorithms becomes vital. This paper sets out to meet this important need.

It is essential to first refer to previous overviews of quantum algorithms: Montanaro's  \cite{Montanaro_2016}, which provides a broad examination of quantum algorithms, emphasizing their main applications and their precise performance bounds; the ``Quantum Algorithm Zoo'' \cite{SJordan}, which serves as a wide repository of quantum algorithms, including 435 references (on 6 April 2024); Mosca's~\cite{mosca2008quantum}, which focuses on quantum computational models and complexity; and several others \cite{Childs_2010, 10.1007/978-3-540-79228-4_3, Zhang_2022}.
%presenting quantum algorithms for algebraic problems \cite{Childs_2010}, quantum walk-based search algorithms \cite{10.1007/978-3-540-79228-4_3}, and quantum algorithms for NISQ era \cite{Bharti_2022}.
The present work builds upon these overviews, with some important differences. We do not focus on complexity speedups over classical algorithms. Instead, we focus on the classification of the vast array of existing quantum algorithms, according to specific criteria, chosen for their ability to reveal different aspects of the algorithms, namely, the fundamental mathematical problems they solve, their real-world applications, the computational model they use, and several others; this constitutes the first core objective of this work, which we meet by the production of a large classification table, see \mbox{Sec.\ \ref{sec:Methodology}} for the methodology followed and the description of the chosen classification criteria. We also establish a genealogy of quantum algorithms by analyzing their structural dependencies; this allows us to identify both the main, established algorithmic primitives, and some of the newer, emerging primitives; all this constitutes part of the second core objective of this work, second core objective which is the analysis of the classification table we have produced, analysis that we present in Sec.\ \ref{sec:analysis}. Although we made our best attempt to analyze a broad-enough sample of representative, major quantum algorithms, our categorization cannot, of course, be fully exhaustive. We apologize for any major omission we may have done. The number of papers in this area of research is growing faster day by day. Still, our framework could be used over a broader set of quantum algorithms: we make our data and visualization codes available online for anyone willing to undertake such an extension, at  \url{https://github.com/Evi-Ksn/Typology-of-Quantum-Algorithms}.

We believe that this work will be useful in two main ways: firstly, in providing a differential understanding of how the algorithms can be grouped and what are the connections between these groups, and secondly, in providing a practical reference usable in different contexts. For example, industries may use this survey to identify quantum algorithms that are related to their business cases, and to evaluate whether these could be implementable in the Noisy Intermediate-Scale Quantum (NISQ) era. Researchers may find in our survey insights about the mathematical outcomes of different quantum algorithms or about the general trends in the field. They may use our dependency analysis as a roadmap to try and find their way back to the sources of quantum advantage. Lecturers may choose to focus the syllabuses of their courses on quantum algorithms down to the core set of primitives that we identify. Graduate and Ph.D. students may again rely on this roadmap to navigate quantum-algorithms learning routes whilst keeping their preferred applications in sight.

This paper is organized as follows. Sec.\ \ref{sec:Methodology} describes (i) the methodology that we followed, especially in order to  build our large classification table, and (ii) each of the classification criteria that we used. In Sec.\ \ref{sec:analysis}, we present our analysis of this classification table, more precisely, our analysis of the correlations between the different classification criteria. %In Section 4, we will display the various practical applications and utilities of the table. 
We conclude in Sec.\ \ref{sec:conclusion} and briefly mention possible extensions of this work. The appendix provides the classification table.

%\section{Related or previous work}
%\label{sec:name}
%The general objective is to
%The quick brown fox jumps over the lazy dog. The quick brown fox jumps over the lazy dog. The quick brown %fox jumps over the lazy dog. The quick brown fox jumps over the lazy dog. The quick brown fox jumps over %the lazy dog. The quick brown fox jumps over the lazy dog. 

%%%%%%%%%%%%%%%%%%%%%%%%%%%%%%%%%%%%%%%%%%%%%%%%%%%%%  
%\chapter{Performed work}
%\label{cap:work}

%\lipsum[1]

\section{Methodology and criteria to build the classification table}
\label{sec:Methodology}
\noindent
To achieve the two core objectives of this work, namely, (i) the classification of the main existing quantum algorithms and (ii) their analysis, we followed various steps sequentially, which we now explain. To make a long story short: we listed a total of 134 quantum algorithms as rows in a large classification table given in Appendix \ref{app:classification_table} (according to the selection process described below in Subsec.\ \ref{subsec:selection_process}); we then listed as columns of this classification table our chosen key criteria, whose description and relevance is given below in Subsec.\ \ref{subsec:classifying_criteria}; the filled classification table finally served as the raw data for our analysis of \mbox{Sec.\ \ref{sec:analysis}}, whose aim was the discovery of relationships, dependencies or patterns among the algorithms and the criteria used, which was achieved by employing statistical and visualization methods.

\subsection{Algorithm selection process}
\label{subsec:selection_process}
\noindent
The selection of the algorithms was by no means random, it was the result of a literature review-based approach, governed by the goal to create an all-inclusive overview that captures the overall landscape of quantum algorithms. The algorithm selection process of course prioritized those algorithms that are highly recognized within the quantum-computing community, ensuring the inclusion of foundational algorithms that have significantly contributed to the advancement of the field over the years. In addition to these most well-known algorithms, one of the several criteria for choosing new ones was their popularity, based on citation frequency. Subsequently, the classification table was updated to try and capture recent developments within current research trends. Digging into each of these topics, we then took the liberty to include further results that appeared to us as quite significant scientifically. Eventually, a last factor that was considered, was the consistency of the classification table itself, e.g., making sure that all main subroutines were present, all application domains inhabited, etc.. 
Again, although we tried to capture the main algorithms using objective and scientific criteria to minimize bias, it is inevitable that some degree of arbitrariness remains in the selection process.

\subsection{Classification criteria}
\label{subsec:classifying_criteria}
\noindent
We now outline the different classification criteria that were chosen in order to explore and enclose the various dimensions of quantum algorithms. 
Examples of algorithms will be given for each criterion.

\noindent
\subsubsection{Main subroutine} %\\[5pt]
%A given algorithm may use one or several \emph{primitive algorithms}, also simply called \emph{primitives}.
%A primitive algorithm is an independent algorithm, typically used in a larger, more complex algorithm, and that does not rely on independent sub-algorithms.
%The term ``routine'' or ``subroutine'' is sometimes used in a similar sense, but the terminology ``primitive algorithm'' emphasizes that it contains part of the essence of the whole algorithm.
%In other words, a primitive algorithm is essential building block of the larger algorithm, while the word "routine" suggests secondary importance.
%
\noindent
In algorithm design, a subroutine is a building block that performs one (or several) specific task(s), packaged as a unit and typically used in larger, more complex algorithm. A subroutine can be called every time that this task needs to be performed, and furthermore, each subroutine can call another subroutine, creating a chain of calls. In the context of this work, the ``Main subroutine" criterion, refers to the first layer of subroutines that an algorithm can have, i.e., not including subroutines of subroutines.
In our classification table this criterion can take as ``value'' either some other algorithms that we have listed in our classification table, or the term ``Primitive''. In the latter case, this means that the algorithm is a fundamental building block, i.e., it is not using any other algorithm as subroutine. Furthermore, in a few cases the algorithm is classified as either a ``variant of'' or ``inspired by'' some other algorithm. In the former, the algorithm is a modified or adapted version, while in the latter, it takes inspiration or concepts from another algorithm, using distinct features.
%
%While the area of quantum algorithms was initially led by seminal algorithms like Shor's and Grover, it has now thrived with abundance of novel algorithms. The importance of the criterion of the primitive algorithm, it's not only to classify the known quantum algorithms based on their dependencies on these essential building blocks, but also to examine if many of the new results in quantum algorithms field are just combinations of a few primitives, or whether they are fundamentally novel.
\\\\
\emph{Example:} Shor's algorithm uses as main subroutine the Quantum Phase Estimation (QPE) algorithm.
\subsubsection{Fundamental mathematical problem} %\\[5pt]
\noindent
This criterion refers to the specific mathematical problem that the algorithm is designed to solve.  In Subsec.\ \ref{subsec:extracting} of the analysis, we extracted wider ``classes of mathematical problems'', or ``mathematical classes'', to which these specific mathematical problems belong. The wider class is mentioned in the same criterion column, in each box, before the colon sign ``:''.\\

\noindent
\emph{Example:} In the case of Shor's algorithm again, the fundamental mathematical problems solved are prime-numbers factorization and computing discrete logarithms, which belong to the wider class of ``Hidden-Subgroup Problems'', see Subsec.\ \ref{subsec:extracting} for the description of this class.

\subsubsection{Applications} %\\[5pt]
\noindent
This criterion refers to the particular problem-solving context(s) in which the algorithm can be useful. Just as for the ``Fundamental mathematical problem'' criterion, we first listed the possible applications of each algorithm. Then, in Subsec.\ \ref{subsec:extracting}, we extracted the wider ``domains of application'', or ``application domains'', that encompass them. The wider domain is mentioned in the same criterion column, in each box, before the colon sign ``:''.\\

\noindent
\emph{Example:} The quantum counterpart of the Principal-Component Analysis, has applications in the domain ``Machine Learning \& Data Science'', see Subsec.\ \ref{subsec:extracting} for the description of this application domain.

\subsubsection{Heuristic versus Proven algorithms}
\noindent
A \emph{heuristic algorithm}, or \emph{heuristic}, is a method of solving problems that uses practical and intuitive techniques to quickly find an \emph{approximate} solution. Usually, this is because the problem is too difficult or impossible to solve exactly in an optimal way.
Heuristics are often based on past experience or rules of thumb, and their solutions come without guarantee.
The evaluation of the performance of these algorithms is empirical and statistical.

Unlike exact, i.e., proven quantum algorithms such as Shor's or Grover's, which have been proven mathematically to provide complexity speedups over the best known classical algorithms for certain problems, heuristic quantum algorithms do not come with a formal proof of their performance. 
Instead, these algorithms are based on intuition and experimentation, and are often motivated by insights from classical algorithms or by physical principles. Indeed, even by these standards, the term ``heuristic'' is perhaps something of a misnomer, as in many cases there is a paucity of \textit{empirical} evidence that the quantum algorithms in question will deliver useful quantum advantage at scale. Nevertheless, the ``heuristics'' that we included in our classification table constitute important algorithms because of their use in experiments probing the capabilities of early NISQ devices.
\\\\
\emph{Example:} The Quantum Approximate Optimization Algorithm (QAOA) \cite{Zhou_2020} is at the heart of numerous quantum heuristics.

\subsubsection{NISQ versus LSQ algorithms}
\noindent
A Large-Scale Quantum (LSQ) computer is a theoretical type of quantum computer that can perform a wide range of tasks and applications beyond those that can be performed by current or near-term quantum computers. It is typically characterized by having a large number of qubits (of the order of millions or more) -- notably, sufficiently many so that effective error correction is possible -- and high-fidelity gate operations. The development of a large-scale general-purpose quantum computer is considered to be a major milestone in the field of quantum computing, as it is expected to enable significant advances in a wide range of fields, such as cryptanalysis, physics, chemistry, materials science, machine learning, data analysis, or optimization. However, building such a device remains a significant challenge due to the inherent fragility of quantum systems and the difficulty in scaling up quantum processors while maintaining the required level of coherence and control.

The quantum devices that are available commercially or in research laboratories today are more limited and belong, for now at least, to the so-called Noisy Intermediate-Scale Quantum~(NISQ) regime. NISQ devices are a necessary step on the path towards LSQ computers, but in principle they could already enable quantum computational advantages, at least in specialized problems that have been explicitly designed to demonstrate a separation between classical and quantum performances. The original definition of NISQ devices is that they contain 50 to 100 qubits and that they are sensitive to the noise induced by their environment, the so-called quantum decoherence \cite{Preskill2018quantumcomputingin}. They suffer from errors, and have insufficiently many qubits to perform effective error correction.

In the context of the classification table, the algorithms were hence separated as either of the NISQ type or of the LSQ type, according to their design, indicating whether they are intended for quantum processors in the NISQ or the LSQ eras. Whilst NISQ versus LSQ is a useful division for the purpose of algorithm classification, which is the intention of this paper, more generally it is worth appreciating that this is somewhat of a false dichotomy insofar as that the first generations of error-corrected quantum computers will still be heavily resource-constrained, and hence NISQ-like design principles (e.g., decomposing circuits to use as few qubits and gates as possible, handing off some computations to classical computers where possible, etc.) will still have to be employed to achieve effective computation for real-world problems. \\
%\todo{Put the links to all the reported experiments in NISQ algorithms (there are missing links)}

\noindent
\emph{Example:} Both the QAOA, for approximate solutions to combinatorial optimization problems, and the Variational Quantum Eigensolver (VQE) for solving quantum chemistry problems, can be classified as NISQ-era algorithms.
\subsubsection{Oracular (versus Non-oracular) algorithms} %\\[5pt]
\noindent
An oracular algorithm is a type of algorithm that makes use of a call to an unspecified generic function, also known as (a.k.a.) ``oracle'', to solve a specific computational problem. Quite often, the problem relates to a property of the given function.

Typically, in the context of quantum computing, the oracle function needs to be made unitary. Thus, the quantum circuit of an oracular quantum algorithm will contain a unitary gate, a.k.a. ``black box'', whose definition depends on the specific oracle that is used. Each use of the black box is referred to as a ``query". In this context, algorithmic complexity is measured in terms of the number of these queries, hence the terminology ``query complexity'' \cite{mosca2008quantum}. %Black-box problems have part of the computational problem hidden in a black-box that must be queried to find a solution.
The strength of oracular algorithms lies in their generality. Their weakness lies in the fact that query complexity ignores many details, including potential simplifications due to specific oracles. 
%On the other hand, a non-oracular algorithm will not use an oracle.
\\\\
\emph{Example:} Grover's algorithm \cite{grover1996fast} for searching for a specific item in an unsorted database of $N$ items, represents that database by means of a black bock $U_f$ that maps the basis state $|x\rangle$ onto $(-1)^{f(x)}|x\rangle$, where $f$ is the oracle Boolean function telling whether $x$ is the desired item.
%\\\\
%More formally, the oracle function $U_f$ is a unitary operator that acts on the quantum state $|x\rangle$ and performs the following transformation:

%$$U_f|x\rangle = (-1)^{f(x)}|x\rangle$$

%where $f(x)$ is a Boolean function : 
%$$f(x) = \begin{cases} 1, & \text{if } x \text{ is a desired state} \\ 0, & \text{otherwise} \end{cases}$$
%%
%\\\\
%
%\newpage
%\noindent
\subsubsection{Promise algorithms} %\\[5pt]
\noindent
An algorithm is of the ``promise'' type when the input of the algorithm needs, for it to work, to take as input one belonging to a subset of the natural set of all possible inputs. This subset is defined as that for which the input satisfies some strong conditions (or promises) that make the problem well-defined and simpler to solve (or even, simply, solvable with that algorithm). \\

\noindent
\emph{Example:} Simon's algorithm \cite{SimonsAlgo} solves Simon's problem, defined as follows: given a function \( f : \{0, 1\}^n \rightarrow \{0, 1\}^n \) with the promise that for some \( s \in \{0, 1\}^n \), it holds that \( f(x) = f(y) \) if and only if \(x \oplus y \in \{\{0\}^n, s\} \), find the \( n \)-bit string \( s \), by making as few queries to \(f(x)\) as possible.
%An example of a promise quantum algorithm is the HHL (standing for Harrow, Hassidim and Lloyd) \cite{Harrow_2009} for solving linear equations. Consider an \textit{N}x\textit{N} matrix \textbf{A} and a vector $\textbf{c} \in \mathbb{R}^N$, the problem is to find a \textit{x} , such that \textbf{A}\textit{x}=\textbf{c}. The HHL algorithm solves efficiently the problem only if the matrix \textbf{A} satisfies the condition (1) to be "sparse", which means that each row of the matrix should consist of at most \textit{y} element and \textit{y}$\ll$\textit{N} and (2) the condition number $\textit{k} = ||A^{-1}||\ ||A||$ which is a parameter measuring the numerical instability of \textbf{A}.
%So the input matrix should satisfy these two conditions in order for the HHL to solve the problem.

\noindent
\subsubsection{Sampling algorithms} %\\[5pt]
\noindent
In the context of computation, sampling problems refer to the task of generating samples from a given probability distribution. The probability distribution may be defined in purely classical terms, or be defined as arising from some underlying quantum process. For example, suppose we prepare an $n$-qubit initial state $|0\rangle^{\otimes n}$ and we apply a series of quantum gates, e.g., a quantum circuit implementing some unitary $U$, before we measure it all in the computational basis. This results in a final state, i.e., an $n$-bit string $x \in \{0, 1\}^n$ sampled with probability $p_x = |\langle x |U|0\rangle^{\otimes n}|^2$~\cite{Lund_2017}. One sampling problem is to produce $x$ with probability $p_x$. \\
%\textcolor{blue}{As special cases, some quantum algorithms could be of "wave function" type. This means the purpose of the algorithm is to prepare a wave function instead of extracting information from it: every step of algorithm is reversible, and the algorithm do not require any post-processing step e.g. measurement.} 
%

\noindent
\emph{Example:} The Boson-Sampling (BS) problem, introduced by Aaronson and Arkipov \cite{aaronson2010computational}, has the following setup: $n$ non-interacting identical photons pass through a linear optical circuit, which is used as an interferometer. The task is to sample the probability distribution of their output positions. This model is easily simulated on a standard LSQ computer, whereas efficient classical BS would imply a polynomial hierarchy collapse.

\noindent
\subsubsection{Partial dequantization} %\\[5pt]
\noindent
A recent \cite{TangDequant}, major realization has been that numerous quantum algorithms can be classically simulated efficiently in some limited regimes (e.g., low rank) by means of classical sampling mimicking the process of state preparation. We have identified which of the algorithms we listed have given rise to quantum-inspired classical algorithms using this methodology. We refer to this criterion as \emph{partial} dequantization, because typically the classical algorithm will serve as a \emph{partial} replacement only (of its quantum counterpart), i.e., it will be applicable to a subset only of possible inputs.

\noindent\subsubsection{Computational model} %\\[5pt]
\noindent
Generally, the expression ``model of computation'', or ``computational model'', in computability an theory, refers to a formal mathematical definition of ``what a computer is''. It thus provides a theoretical framework about how some function may be computed, by specifying things such as: the presentation/encoding of inputs and outputs, the computational steps, the memory structure, and the communication methods. The computational complexity of an algorithm can then be defined in those terms \cite{modelsofcomp}. 

In quantum computing, there exists a variety of computational models. We will list the most important ones, and mark them with the sign ``($\ast$)'' when they have been used in our typology, i.e., as a possible ``value'' taken by the criterion ``Computational model'' in the relevant column of our classification table. We choose not to abbreviate or use acronyms for the computational models in this work. Among these computational models we find, most relevantly, Quantum Turing Machines \cite{qtm_paper,universal_qtm} -- often acronymized QTMs in the literature --, the Circuit Model~($\ast$)~\cite{circuit_model}, Quantum Cellular Automata ($\ast$) \cite{qca_review}  -- often acronymized QCAs in the literature --, Adiabatic Quantum Computation ($\ast$) \cite{adiabatic_review} -- often acronymized AQC in the literature --, Quantum Walks ($\ast$) \cite{qw_review} -- often acronymized QWs in the literature --, or Measurement-Based Quantum Computation ($\ast$) \cite{mbqc_review} -- often acronymized MBQC in the literature. All of these computational models can do universal quantum computation in the sense of approximating any arbitrary unitary operation $U\in \text{U}(n)$, $n \in \mathbb N$, up to arbitrary precision \cite{ArrighiUniversalQCA, mbqc_universal, qw_universal, adiabatic_universal}.

Note that some of these models can be viewed as mathematical abstractions of certain physical implementations or experiment protocols: Quantum Annealing approximately implements  Adiabatic  Quantum Computation \cite{annealing_review}, Linear Optical Quantum Computation ($\ast$) -- often acronymized LOQC in the literature -- is the preferred architecture to realize Measurement-Based Quantum Computation \cite{loqc_1, loqc_2}, some  architectures for the Circuit Model correspond to Quantum Random-Access Memories ($\ast$) \cite{qram_review}  -- often acronymized QRAMs in the literature --, etc..

\subsubsection{Citations} %\\[5pt]

\noindent
The last criterion column of the classification table is that of the number of Google-Scholar citations. For each algorithm of the classification table, we picked as reference either the seminal paper explaining this algorithm, or a very relevant paper explaining this algorithm, and then consulted the number of citations of this reference paper, on Google Scholar. This check was done between October 2023 and June 2024. We have not used this number to weight our data visualization. We used it to determine (amongst other factors) whether some non-recent papers ought to be present in the classification table or not.

\section{Analysis of the classification table}
\label{sec:analysis}
\noindent
The diagrams presented in this part can be found at \url{https://github.com/Evi-Ksn/Typology-of-Quantum-Algorithms} for a better resolution.

\subsection{Dependency and algorithmic-primitives analyses}\label{subsec:dependency}
\noindent
Whilst the area of quantum algorithms was, two or three decades ago, dominated by a handful of seminal algorithms like Shor's and Grover's, it has now expanded into a rich landscape of numerous novel algorithms. However, a legitimate question is whether the new algorithms of this field are made from combinations of the old ones used as subroutines, or whether they are fundamentally novel altogether. Analyzing the ``Main subroutine'' criterion of the classification table, can give us an answer to this question.

\subsubsection{Dependency network}

\begin{figure}[h]
  \vspace{-2cm}
  \hspace{-2cm}
  \rotatebox{-90}{\includegraphics[height=20cm]{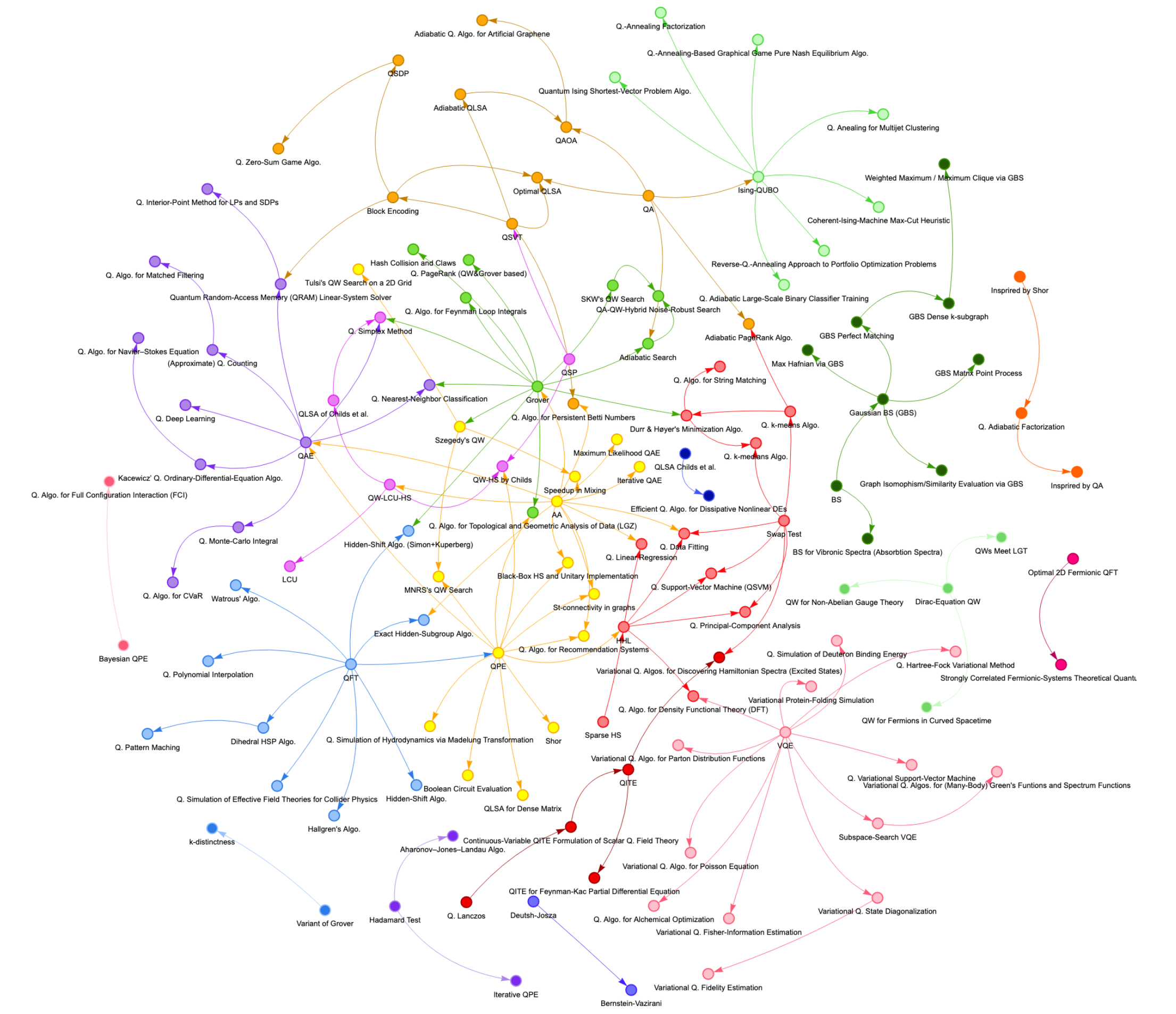}}
  \caption{{\bfseries Dependency Network.}}
  \label{fig:network}
\end{figure}

\noindent
First of all, we used these data of the column ``Main subroutine'' to generate a network of dependencies (or dependency network) of quantum algorithms, that we present in Fig.\ \ref{fig:network}. Let us first describe very generally the dependency network. It provides a detailed visualization of all the layers of subroutines that are used as ``steps'' in each algorithm. It offers an insight into the logical relationships between these algorithms, revealing also, to some extent, the genealogy of ideas underlying their design.

Let us now describe the dependency network in more detail. In this dependency network, arrows go from some subroutine to the algorithm making use of that subroutine, subroutine which may itself use another subroutine, and/or be used by another algorithm. An algorithm sometimes uses several subroutines. Primitive algorithms, are thus shown as root nodes, i.e., without any arrow pointing towards them. When a primitive is not used by any other algorithm, it does not appear in our dependency network. When we write ``a variant of'' or ``inspired by'' some algorithm in the classification table, that latter algorithm is treated in the dependency network exactly as if it was a subroutine -- i.e., there's an arrow from it towards the algorithm indicated as ``a variant of'' or as ``inspired by''. Notice that some dependencies vary with the version of the algorithm: for instance,  whilst the original QPE relies on the Quantum Fourier Transform (QFT), there exists Bayesian QPE or Iterative QPE, which do not; similarly, different Quantum Amplitude Estimation (QAE) algorithms may rely on QPE or not. Notice also that several algorithms dependent on some early Quantum Linear-System Algorithm (QLSA), could in fact be made dependent on the Optimal QLSA~\cite{costaQLSA}.

Let us now start analyzing the dependency network. By examining it, we clearly observes four large clusters. At their respective centers lie four core primitives: Quantum Fourier Transform~(QFT)~--~or the original QFT-based QPE, if we allow ourselves to speak about non-primitive subroutines --, Amplitude Amplification (AA), Quantum Adiabatic (QA), and Variational Quantum  Eigensolver (VQE). 
It is interesting, and reassuring, that these four primitives having emerged as ``cores'' through the methodology followed by this study matches with what a practitioner in quantum algorithms would expect. One can also witness on this dependency network, to some extent, the consequences of a natural shift of interest in quantum algorithms (this will be clearer further down). Indeed, let us first notice that two important and distinct sources of quantum advantage that have been identified since the early days of quantum computing are the following: the original, QFT-based QPE, underpinning Shor's algorithm; and AA, underpinning Grover's algorithm.
But, with the advent of real-world quantum hardware, attention has been drawn towards running algorithms on present-day devices seeking to implement VQE and, to some extent, QA; that is to say, since the proliferation of small-scale quantum hardware, the focus has switched from algorithms exhibiting asymptotic quantum advantage, to those that can run on such small-scale quantum hardware (even, on occasions, algorithms for which there is little evidence that there will be quantum advantage as the hardware scales).

\subsubsection{Algorithmic-primitives analysis}

\noindent
Going back to the first column ``Main subroutine'' of our classification table, and focusing on the Primitives, one immediately sees that, beyond 8 usual ``suspects'', many other primitives, or non-primitive popular subroutines (not necessarily used by other algorithms) could be listed. Those findings are summarized in Table \ref{tab:primitives}. 
In the first column of that table, we listed the 8 usual primitives mentioned above. In the second column, we listed other, non-primitive popular subroutines, either already old, or new ones, more or less half of which are already indeed used as subroutines of other algorithms. In the last column, we listed the remaining primitives.

\begin{table}
    \hspace{-0.5cm}
\begin{tabular}{lll}% 其中，tabular是表格内容的环境；c表示centering，即文本格式居中；c的个数代表列的个数
\toprule %[2pt]设置线宽     
\parbox[l][0.4cm][l]{3.3cm}{
\begin{spacing}{0.6} Popular Primitives  \end{spacing}}  & \parbox[l][0.4cm][l]{4.1cm}{
\begin{spacing}{0.6} Popular Subroutines  \end{spacing}}  & \parbox[l][0.4cm][l]{5.4cm}{
\begin{spacing}{0.6} Other Primitives \end{spacing}} 
\\ %换行
\midrule % [2pt]  
\parbox[c][0.4cm][c]{3.3cm}{
\begin{spacing}{0.6} QFT \cite{QFT1}  \end{spacing}}
&     \parbox[c][0.4cm][c]{4.1cm}{
\begin{spacing}{0.6}QPE \cite{qpe_paper}  \end{spacing}}&  \parbox[c][0.4cm][c]{4.5cm}{
\begin{spacing}{0.6} QITE \cite{qite}  \end{spacing}}   

\\
\parbox[c][0.6cm][c]{3.3cm}{
\begin{spacing}{0.6}  AA \cite{aa_original}   \end{spacing}}  & \parbox[c][0.6cm][c]{4.1cm}{
\begin{spacing}{0.6}   QAE \cite{qae_paper}    \end{spacing}}     &  \parbox[c][0.6cm][c]{4.8cm}{
\begin{spacing}{0.6}  Q.\ Lanczos \cite{Motta2019}   \end{spacing}} 

\\  \parbox[c][0.6cm][c]{3.3cm}{
\begin{spacing}{0.6}  QA  \cite{qa_origianl} \end{spacing}}   &   \parbox[c][0.6cm][c]{4.1cm}{
\begin{spacing}{0.6}   Grover \cite{grover_original}      \end{spacing}}   &  \parbox[c][0.6cm][c]{4.7cm}{
\begin{spacing}{0.6}  Variational QSVD \cite{Bravo_Prieto_2020} \end{spacing}} 

\\ \parbox[c][0.6cm][c]{3.3cm}{\begin{spacing}{0.6}  
VQE \cite{vqe} 
\end{spacing}
}  
  & \parbox[c][0.6cm][c]{4.1cm}{\begin{spacing}{0.6}  
QLSAs \cite{Chakraborty2018ThePO,Childs_qsla,costaQLSA,Harrow_2009,PhysRevLett.120.050502}
\end{spacing}
}      &  \parbox[c][0.6cm][c]{4.7cm}{\begin{spacing}{0.6}  
Bayesian QPE  \cite{smith2023adaptive} 
\end{spacing}
}

  \\ \parbox[c][0.6cm][c]{3.3cm}{
\begin{spacing}{0.6}
HSs \cite{aa_la_hs_1, aa_la_hs_2,Berry_2006,BE_paper,Childs:2010,Childs:2012fnt}
\end{spacing}
}
& \parbox[c][0.6cm][c]{4.1cm}{
\begin{spacing}{0.6} QSVT \cite{10.1145/3313276.3316366}    \end{spacing}} &  \parbox[c][0.6cm][c]{4.8cm}{
\begin{spacing}{0.6}  Hadamard Test \cite{aharonov2006polynomial}   \end{spacing}}  

\\

\parbox[c][0.6cm][c]{3.3cm}{
\begin{spacing}{0.6}  Block Encoding \cite{BE_paper}  \end{spacing}}  
 & 
\parbox[c][0.6cm][c]{4.1cm}{
\begin{spacing}{0.6}  QSP \cite{qsp}  \end{spacing}}    &  
\parbox[c][0.6cm][c]{5.2cm}{
\begin{spacing}{0.6}  Hilbert-Schmidt Test \cite{Khatri_2019}  \end{spacing}} 

\\
\parbox[c][0.6cm][c]{3.3cm}{
\begin{spacing}{0.6}  BS \cite{Crespi_2013} \end{spacing}} 

 &   
   \parbox[c][0.6cm][c]{4.1cm}{
\begin{spacing}{0.6}   Szegedy's QW \cite{aa_combi_search_segzdy}   \end{spacing}}  
 &  
\parbox[c][0.6cm][c]{5.4cm}{
\begin{spacing}{0.6}  Q.\ Hierarchical Clustering \cite{patil2023measurementbased}  \end{spacing}}

\\
\parbox[c][0.6cm][c]{3.3cm}{
\begin{spacing}{0.6}   Swap Test \cite{barenco1996stabilisation}  \end{spacing}}  
   &  
\parbox[c][0.6cm][c]{4.3cm}{
\begin{spacing}{0.6}  SKW's QW Search \cite{search_QW1}    \end{spacing}} 
  &    
 \parbox[c][0.6cm][c]{4.8cm}{
\begin{spacing}{0.6}   Q.\ Distance Classifier \cite{Schuld_2017} \end{spacing}}

 \\
 \parbox[c][0.6cm][c]{3.3cm}{
\begin{spacing}{0.6}    \end{spacing}}  
              & 
              \parbox[c][0.6cm][c]{4.1cm}{
\begin{spacing}{0.6}   Adiabatic Search \cite{search_adiabatic}    \end{spacing}}  
                &  \parbox[c][0.6cm][c]{7.6cm}{
\begin{spacing}{0.6}   Q.-Enhanced Markov-Chain Monte Carlo \cite{Layden_2023}  \end{spacing}}    
              
              \\
               \parbox[c][0.6cm][c]{3.3cm}{
\begin{spacing}{0.6}    \end{spacing}}  
      &   \parbox[c][0.6cm][c]{4.7cm}{
\begin{spacing}{0.6}    D.\ \& H.\ Minimization \cite{aa_combi_opt_DHminim}    \end{spacing}}    &  \parbox[c][0.6cm][c]{4.8cm}{
\begin{spacing}{0.6}  Simon's Algo.\ \cite{Kaplan_2016}   \end{spacing}}    
      
    \\\parbox[c][0.6cm][c]{3.3cm}{
\begin{spacing}{0.6}    \end{spacing}}  
            & \parbox[c][0.6cm][c]{4.1cm}{
\begin{spacing}{0.6}  Q.\ Counting     \cite{Brassard_1998}      \end{spacing}}    &
             \parbox[c][0.6cm][c]{6cm}{
\begin{spacing}{0.6} Fermionic QFT \cite{Jiang_2018}   \end{spacing}}           
      
      \\\parbox[c][0.6cm][c]{3.3cm}{
\begin{spacing}{0.6}    \end{spacing}}  
    &  \parbox[c][0.6cm][c]{4.7cm}{
\begin{spacing}{0.6}    Q.\ Monte-Carlo Integral \cite{QAE_statistics_QMCI}     \end{spacing}}  &
\parbox[c][0.6cm][c]{5.7cm}{
\begin{spacing}{0.6} QCA for Q.\ Electrodynamics \cite{eon2023relativistic}   \end{spacing}}   
    \\\parbox[c][0.6cm][c]{4.8cm}{
\begin{spacing}{0.6}    \end{spacing}}  
            &  \parbox[c][0.6cm][c]{4.1cm}{
\begin{spacing}{0.6}  Gaussian BS \cite{GBS_prl}   \end{spacing}}    &   \parbox[c][0.6cm][c]{4.8cm}{
\begin{spacing}{0.6}    Q.\ Final State Shower \cite{Nachman_2021} \end{spacing}}

            \\\parbox[c][0.6cm][c]{3.3cm}{
\begin{spacing}{0.6}    \end{spacing}}  
              & \parbox[c][0.6cm][c]{4.1cm}{
\begin{spacing}{0.6} Ising-QUBO \cite{ising_qubo}     \end{spacing}  }  &  \parbox[c][0.6cm][c]{7cm}{
\begin{spacing}{0.6} Free-Q.-Field-Theory Ground State \cite{Bagherimehrab_2022}   \end{spacing}}  
              \\\parbox[c][0.6cm][c]{3.3cm}{
\begin{spacing}{0.6}    \end{spacing}}  
 &  \parbox[c][0.6cm][c]{4.1cm}{
\begin{spacing}{0.6}   QAOA \cite{qaoa}   \end{spacing}}     &   \parbox[c][0.6cm][c]{6cm}{
\begin{spacing}{0.6}  Q.\ Fock Space Dynamics \cite{Yao_2023}  \end{spacing}}

 \\\parbox[c][0.6cm][c]{3.3cm}{
\begin{spacing}{0.6}    \end{spacing}}  
  &  \parbox[c][0.6cm][c]{4.1cm}{
\begin{spacing}{0.6} QSDP \cite{q_enable_qsdp}     \end{spacing}}   &   \parbox[c][0.6cm][c]{6cm}{
\begin{spacing}{0.6} Dirac-Equation QW \cite{BialynickiBirula1994}   \end{spacing}}     
  \\\parbox[c][0.6cm][c]{3.3cm}{
\begin{spacing}{0.6}    \end{spacing}}  
    & \parbox[c][0.6cm][c]{4.7cm}{
\begin{spacing}{0.6}    Dihedral HSP Algo.\ \cite{kuperberg2004subexponentialtime} \end{spacing}}   &  
    
  \\\parbox[c][0.6cm][c]{3.3cm}{
\begin{spacing}{0.6}    \end{spacing}}  
    & \parbox[c][0.6cm][c]{4.7cm}{
\begin{spacing}{0.6}    Shor \cite{Shor_1997} \end{spacing}}   &

                        \\
\bottomrule %[2pt]   
\end{tabular}
\caption{Table of primitives and non-primitive popular subroutines (none of the subroutines of the second column is, indeed, a primitive). Note that in order to be intuitive, all HS algorithms are clustered as ``HSs'', and all QLSA algorithms are clustered as ``QLSAs''. The same convention applies for Fig.\ \ref{fig:matrix}, which uses the second column only, and for Fig.\ \ref{fig:correlation}, which uses the first column only. }
\label{tab:primitives}
\vspace{1cm}
\end{table}

Let us briefly speculate about which primitives or popular subroutines may be most dominant in the near or further future. It seems likely that the Quantum Singular-Value Transformation~(QSVT) will play a prominent role in the future. Indeed, QSVT is famously a \textit{unifying} algorithm, that subsumes most of the other prominent algorithms as special cases \cite{GilyenUnifying,MartynUnifying}. Moreover, QSVT also plays the role of a \textit{framework} that captures the ``modern'' approach to quantum-algorithm design, namely, using quantum computers to manipulate the eigenvalues or singular values of exponentially large matrices -- applied to some suitable input. This idea is extremely powerful, and the first hint of it actually came in the Harrow-Hassidim-Lloyd (HHL) algorithm, for which one of the primitives is QPE. Indeed, unlike most other algorithms that use QPE, rather than just extracting a particular eigenvalue phase, HHL manipulates the quantum state by accessing the several phases of eigenvalues in superposition in order to reach the desired result.  

An interesting development is Tang's \cite{TangDequant} methodology for extracting quantum-inspired classical algorithms from existing quantum algorithms, which are efficient for a certain range of parameters. This methodology is useful mainly for the HHL, QPE, and QLSA algorithms (a.k.a. qBLAS algorithms \cite{Biamonte_2017}), with impact on linear-algebra problems and machine-learning and data-science applications, amongst others. But the methodology can also be used for QSVT. Again, given the unifying nature of the latter, it is at this stage rather difficult to gauge the potential ramifications of dequantization that may ensue from a use of Tang's methodology on QSVT.

\clearpage
\subsection{Extracting mathematical classes and application domains}
\label{subsec:extracting}
\noindent 
%
%In this subsection, we will use the denomination “Mathematical Problem” to designate what we have called above “Fundamental Mathematical Problem”, just for compactness of the wording.
%
After thoroughly filling, for each algorithm, the specifics (i) for the ``Fundamental mathematical problem'' criterion/column of our classification table, and (ii) for the ``Applications'' criterion/column, we realized the need to take a step back to get to the bigger picture.
\\\\
For the ``Fundamental mathematical problem'' criterion, we identified 6 classes, described below.

\begin{itemize}
  \item \textbf{Hidden-Subgroup Problems:} This class of problems is about identifying a subgroup from a function that is constant ``on'' each coset of that subgroup and that takes distinct values on different cosets of this subgroup. Let us give a formal definition. First of all, let us define what a coset of a subgroup is. A left coset of some subgroup $S$ of $G$ is a set $gS$ for some $g \in G$. The definition for right cosets is analogous. A fundamental result about cosets is that two left cosets or two right cosets are either the same or disjoint. Let us now define the standard Hidden-Subgroup Problem (HSP). Given a group $G$, a finite set $X$ and a function $f : G \rightarrow X$, the goal is to identify a hidden subgroup $H \subsetneqq G$ of the group $G$ such that $f(x)$ is constant ``on'' each coset of $H$, which means that, for all $g_1, g_2 \in G$, $f(g_1) = f(g_2)$ if and only if $g_1H = g_2H$ \cite{lomont2004hidden}. Shor's algorithm, which solves the problems of prime-numbers factorization and of computing discrete logarithms, can be generalized as an HSP for finite Abelian groups~\cite{van_Dam_2012}.
  %This is computing a subgroup from a function that is constant on the cosets of the subgroup and takes distinct values on different cosets see the discussion of the main result for a formal definition.
  %
  \item \textbf{Linear Algebra:} Linear-algebra problems deal with linear equations, linear transformations and their representations using matrices. For example, the Quantum Singular-Value Decomposition (QSVD) algorithm \cite{Bravo_Prieto_2020} produces the singular-value decomposition of a bipartite pure state, which is indeed a linear-algebra problem.
  \item \textbf{Dynamical Systems:} A dynamical system is a system whose state changes over time. This could be either (i) in discrete time, via finite-difference equations in either only time, or both time and space, or (ii) in continuous time, with a partial derivative with respect to time at least, and, if we are also in continuous space, another partial derivative, with respect to space this time, leading to a partial differential equation (otherwise it is an ordinary differential equation with respect to time, with finite differences in space). Quantum algorithms that belong to this class seek to predict and analyze the system's time evolution, e.g., by preparing an initial state, encoding information therein, evolving it, or evaluating key features of this evolution. For instance, the Quantum Lanczos algorithm~\cite{Motta2019} computes the ground state of some Hamiltonian, which is a key feature of the evolution of the system described by the Hamiltonian, so that this algorithm belongs to the class ``Dynamical Systems''. 
  \item {\textbf{Stochastic Processes \& Statistics:} 
  A stochastic process, also called a random process, is a mathematical concept used to model and describe the evolution of variables/quantities that evolve randomly over time (or some other parameter), using probabilistic rules. %I can cite this https://arxiv.org/pdf/0909.4213.pdf if needed
  Statistics is a branch of mathematics that studies the collection, analysis, interpretation, and presentation of data, so as to make inferences or decisions based on them. 
  %These two concepts are essential for studying uncertainty and randomness in various domains. 
  Typically, the stochastic process is what generates the data that statistics analyze.
  In the context of this work, an example of an algorithm that solves a stochastic-processes problem is the Gaussian-Boson-Sampling (GBS) Matrix-Point Process \cite{Jahangiri_2020} -- which, in addition, is a sampling algorithm, as defined above. An example of an algorithm that solves a statistics problem is the Quantum k-means algorithm \cite{lloyd2013quantum}.
  }
  %
  %\item \textbf{COV:} Convex optimization. Convex optimization problems deal with finding the minimum (or maximum) of a convex function over a convex set. In the "Q. Linear regression" algorithm \cite{Schuld_2016}, the fundamental mathematical problem is linear regression using least squares optimization, which falls into "COV" realm.
  %which one we keep? maybe it should be better to make the seperation
  \item \textbf{Optimization:} ``Optimization'' refers to the class of problems that deal with finding the best solution amongst a given set of candidates. The ``best'' solution is determined by minimizing or maximizing some target function while satisfying a set of constraints. Optimization includes numerical optimization and combinatorial optimization problems. When an algorithm is of the combinatorial-optimization type, we assigned  it both the ``Optimization'' class described here, and the ``Combinatorics'' class described below; otherwise~(algorithm of the numerical-optimization type), we just assigned it the ``Optimization'' class. An example is the Quantum Linear Regression algorithm \cite{Schuld_2016}, whose fundamental mathematical problem is linear regression using least-squares optimization, which is convex optimization (which is a a subclass of numerical optimization).
\item \textbf{Combinatorics:} Combinatorial problems involve finding, for a discrete finite set of objects that satisfy some given conditions, either a grouping, or an ordering, or an assignment. They can be divided into tree basic types, that are, enumeration problems, existence problems, and optimization problems. An algorithm classified under ``Combinatorics'' in the classification table, focuses on solving problems like graph-theoretical problems, search tasks, counting tasks, etc.. For example, Grover's search algorithm has been extended to Quantum Counting \cite{Brassard_1998}, and both can be classified as combinatorial.
\end{itemize}
\newpage
Regarding the ``Applications'' criterion, we extracted 6 application domains, described below.
\begin{itemize}
  \item \textbf{Cryptanalysis:} Cryptanalysis refers to the study of cryptographic systems with the aim of understanding their strengths and weaknesses. Most works in this domain study the applicability of quantum algorithms to break classical cryptographic schemes. The obvious, early example is Shor's factorization algorithm, whose LSQ implementation would break the RSA public-key encryption system \cite{10.1145/357980.358017}.
  
  \item \textbf{Machine Learning \& Data Science:}  This domain involves (i) performing tasks like classification, regression, clustering, etc., and (ii) applications of quantum algorithms in the fields of data analysis, statistical modeling, or predictive analytics for managing large datasets. For instance, the Quantum Algorithm for Recommendation Systems \cite{kerenidis2016quantum} has applications in both these areas.
  
  \item \textbf{First-Principle Quantum Simulation:} ``First-Principle (also called ``ab initio'') Quantum Qimulation'' refers to using quantum computing devices in order to simulate and predict physical properties of a system, based upon its fundamental quantum mechanical description -- without any detour via empirical models or parameter fittings. Indeed, although classical algorithms for matrix arithmetics have only polynomial complexity in the matrix dimension, the Hilbert-space dimension of multi-particle systems/many-body problems is exponential in the system's size. Thus, quantum algorithms are bound to quickly overcome their classical competitors \cite{mcclean_exploiting_2014}. In fact, using programmable quantum systems to simulate quantum physical systems was Feynman's original motivation for inventing the very concept of a quantum computer \cite{feynman_simulating_1982}. Typically, first-principle quantum simulations involve solving particle-structure problems, electronic structures \cite{HF2020}, atomic-nucleus structures~\cite{AtomicNucDMH2018}, or material properties. We distinguish this application domain from other scientific computing approaches, which we refer to as ``classical'' below.
  
\item \textbf{Classical Scientific Computing:} This domain is used to accommodate scientific-computing applications that are not first-principle quantum simulations, including empirical-modeling applications and data/signal-processing applications. Quantum algorithms of this application domain may outperform their classical counterparts for specific tasks, including classical-mechanics simulations \cite{NS2022LB}, empirical/semi-empirical numerical model prediction and optimization \cite{GBS1, dissolve2021JPL}, and experimental data processing/mining \cite{Gravity2022GSHF}.
  
\item \textbf{Operational Research:} ``Operational Research'' involves employing analytical methods, like modeling, statistics, optimization, in order to arrive at optimal or near-optimal solutions to decision-making problems, in complex scenarios like logistics, supply-chain management, resource allocation, etc.. Within the context of this work, mathematical programming, optimization, and game-theory algorithms are classified as ``Operational Research''. As an example, let us cite the Quantum Algorithm for CVaR \cite{Woerner_2019}, which has a specific application in finance, for risk management.
  
  \item \textbf{Quantum Enablement:} This domain covers the particular quantum algorithms, closer to the hardware part, that underlie the experimental implementation of other quantum algorithms, closer to the software part, or belong to the construction of the quantum technical stack itself. Typically, they have no direct applications by themselves, although some  may be applied in quantum sensing or metrology \cite{q_enable_q_fedelity,q_enable_qsdp}, or to demonstrate quantum advantage \cite{q_enable_bosonsamp,q_enable_bv,q_enable_dj}. They often represent general algorithmic-design ideas on which many more specific quantum algorithms are based \cite{q_enbale_aaa,QFT1}, e.g., quantum test algorithms are classified in this domain \cite{q_enable_hadamard_test, q_enable_swap_test,q_enable_HS_test}.

\end{itemize}

\subsection{Correlations and trends regarding the three principal criteria, ``Main subroutine'', ``Fundamental mathematical problem'' and ``Applications''}
%\todo[inline]{Correlation analysis between the other criteria -- Promise - Oracular - Heuristic - Sampling--done by Pablo}

\noindent
Having coarse-grained fundamental mathematical problems into classes, and applications into domains, helps identifying correlations and trends.

\subsubsection{Correlations between mathematical classes and application domains, related trends, and correlations with important subroutines}

\begin{figure}[h] 
  \centering       
  %\hspace{-2cm}
  \includegraphics[width=0.9\textwidth]{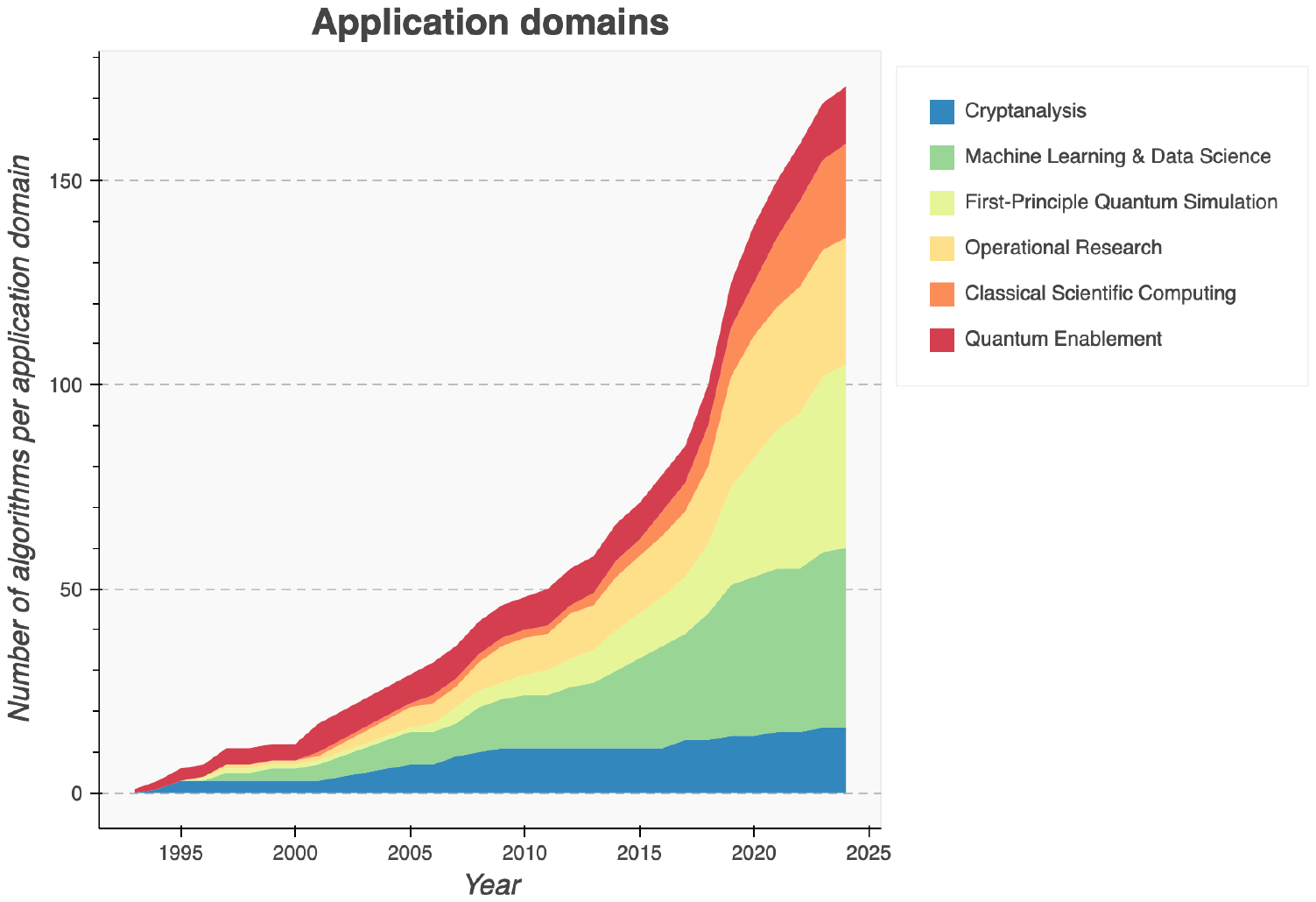} 
  \vspace{-1.5cm}
  \caption{{\bfseries Application-Domains Growth.} Evolution over time of the number of new algorithms per application domain.}
  \label{fig:final_stacked_app}
    \vspace{1cm}
\end{figure}

\noindent
In Fig.\ \ref{fig:final_stacked_app}, we plot the number of algorithms for each of the 6 application domains, over time. Some trends become apparent. Indeed, the application domains in which there is an increasing number of new algorithms in the last years (here we roughly merge the acceleration of this number and its absolute value when we say ``increasing number'', we do not disambiguate between both features at this stage), are, by decreasing increasing number, ``First-Principle Quantum Simulation'', ``Machine Learning \& Data Science'', ``Operational Research'', and, to a lesser extent ``Classical Scientific Computing'' and ``Quantum Enablement''. The ``Cryptanalysis'' domain is the one that evolves slowest in terms of acceleration of the number of new algorithms.

\begin{figure}[h] 
  \centering       
  %\hspace{-2cm}
  \includegraphics[width=0.9\textwidth]{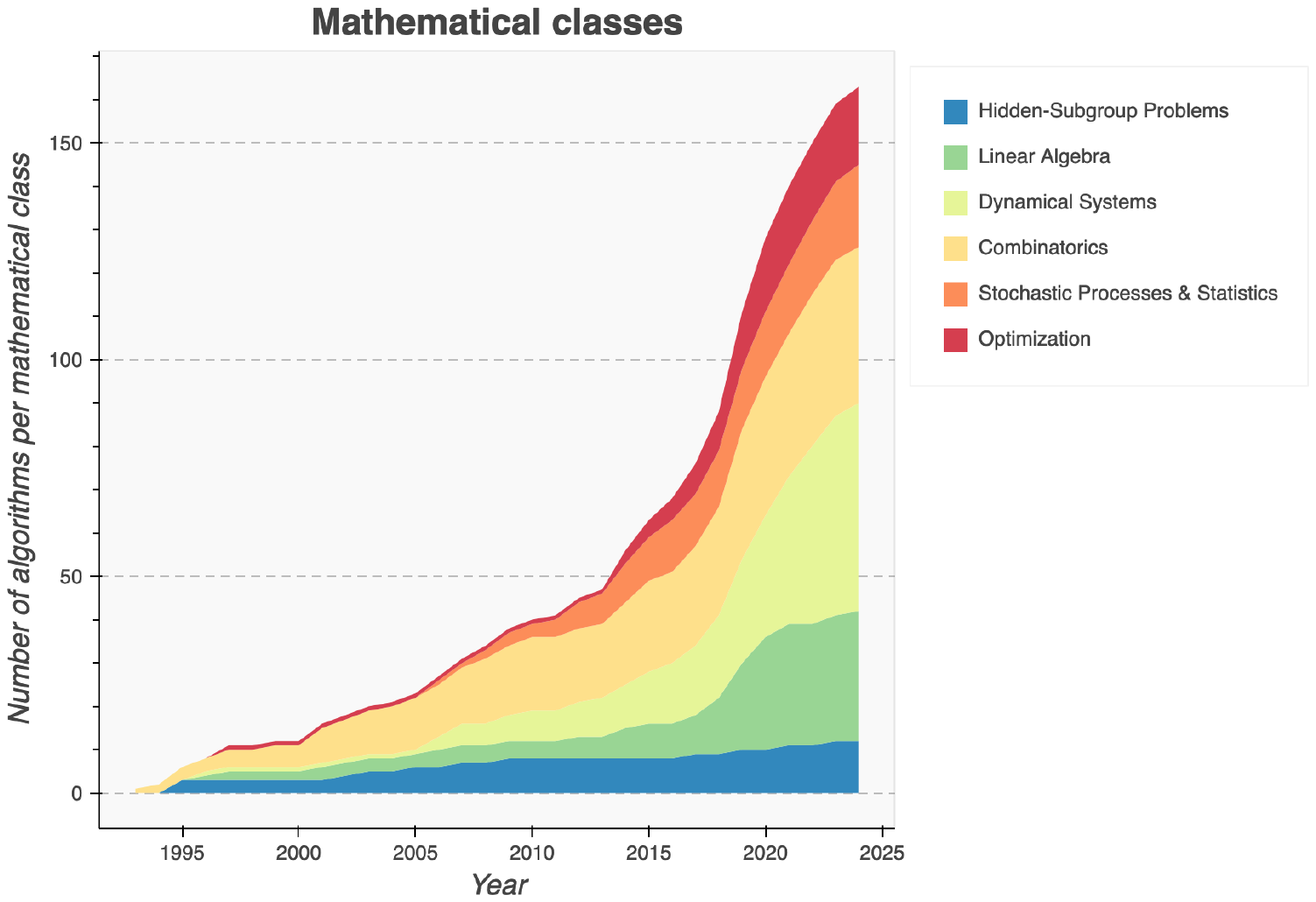} 
  \vspace{-1.5cm}
  \caption{{\bfseries Mathematical-Classes Growth.} Evolution over time of the number of new algorithms per mathematical class.}
  \label{fig:stacked_math_final_paper}
    \vspace{1cm}
\end{figure}

A similar diagram, focusing this time on the mathematical classes, is presented in Fig.\ \ref{fig:stacked_math_final_paper}. It shows that an important proportion of new algorithms fall into the ``Dynamical Systems'' class. This trend aligns with the previously noted growth in ``First-Principle Quantum Simulation'' applications, as these applications often require solving dynamical-system problems. Similarly, an important proportion of new algorithms fall into the ``Linear Algebra'' and ``Optimization'' classes, which again aligns with the previously noted growth in ``Machine Learning \& Data Science'' and ``Operational Research'' applications, since the latter require solving a combination of linear algebra, statistical, and optimization problems. The ``Hidden-Subgroup Problems'' class is stable, correlated with the stability of the ``Cryptanalysis'' application.

The previous trend analyses clearly suggest correlations between mathematical classes and their application domains, through the algorithms serving them. Let us evaluate this more precisely. We have computed these correlations and visualized \cite{raw_graphs} the result in \mbox{Fig.\ \ref{fig:matrix}}. Notice how these findings were enabled by the dependency analysis of Subsec.\ \ref{subsec:dependency}, allowing us to focus on main subroutines for increased robustness: indeed, the x-axes of the two diagrams of Fig.\ \ref{fig:matrix}, which are both the same, correspond to the list of Popular Subroutines identified thanks to the dependency network, and which appears in the second column of Table \ref{tab:primitives}. Hopefully these two diagrams can help trace back the root of some quantum advantage, according to one's objective. 

\subsubsection{Correlations between algorithmic primitives, mathematical classes, and application domains}

\begin{figure}[h!] 
  \centering       
  \includegraphics[width=1\textwidth]{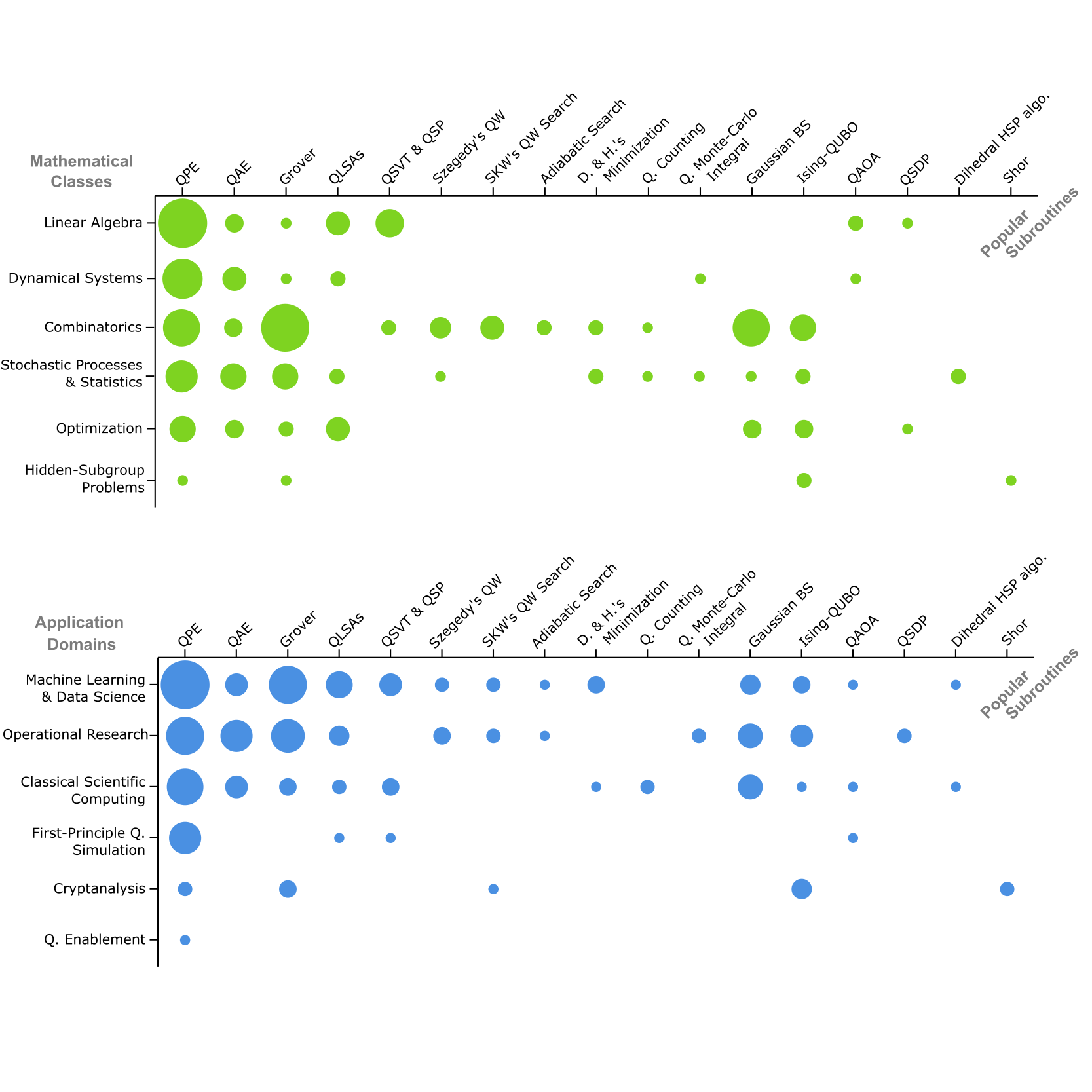} 
  \vspace{-1.5cm}
  \caption{{\bfseries Correlations Array.} Array of correlations between the Popular (non-primitive) Subroutines~(of~the second column of Table \ref{tab:primitives}) and either (i) mathematical classes for the top array, or (ii) application domains for the bottom array.}
  \label{fig:matrix}
   \vspace{1cm}
\end{figure}

\begin{figure}[h!] 
  \centering       
  %\hspace{-2cm}
  \includegraphics[width=1\textwidth]{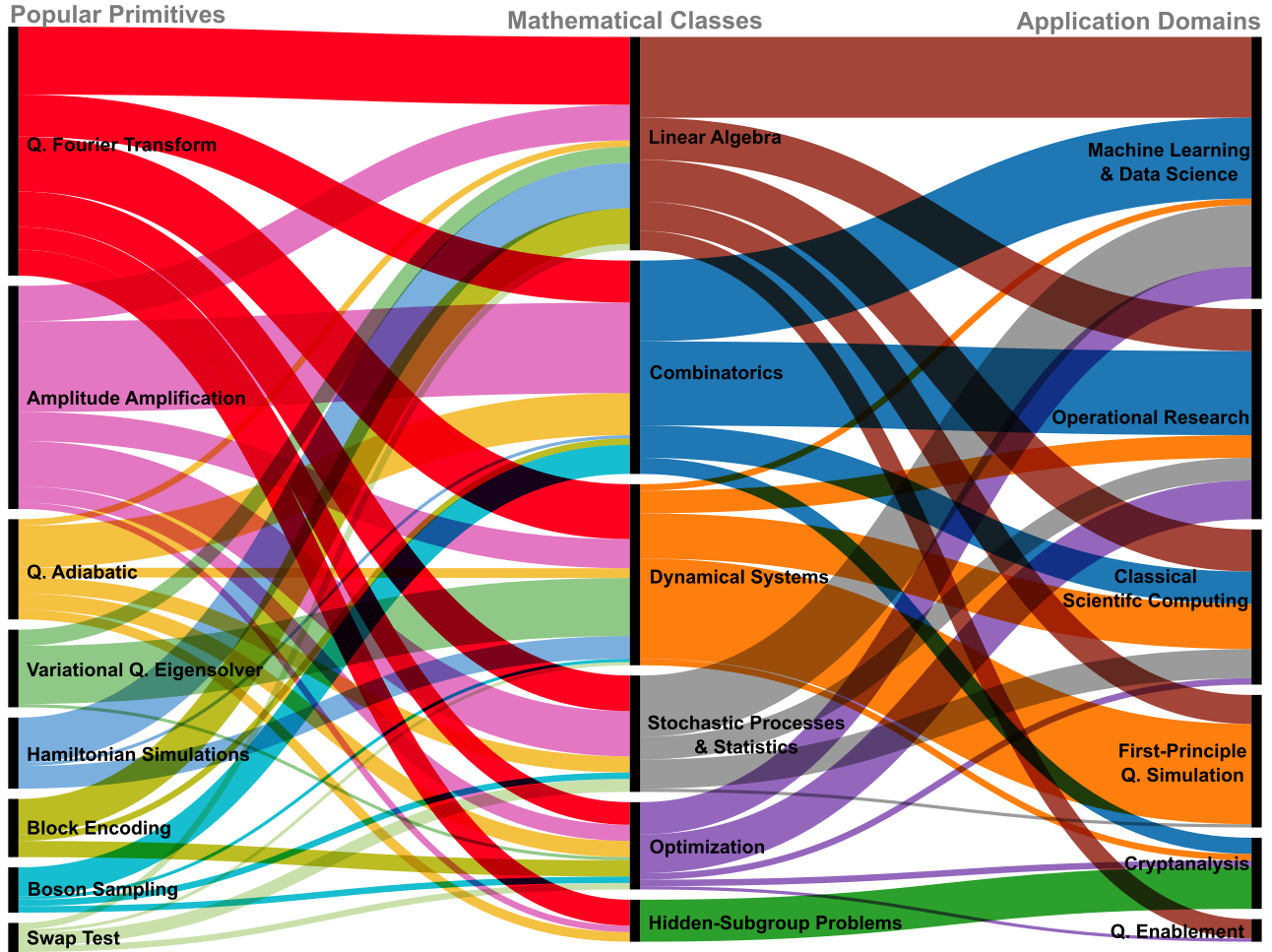} 
  \caption{{\bfseries Sankey Diagram.} Correlation between Popular Primitives (first column of Table~\ref{tab:primitives}), classes of the criterion ``Fundamental mathematical problem'', and domains of the criterion ``Applications''. In this diagram, the meaning of ``Q.\ Adiabatic'' in the column of Popular Primitives is ``any algorithm of the QA type'', i.e., we do not limit ourselves to the original QA algorithm. The same applies to ``Hamiltonian Simulations''. The positions of the ``connections'' which ``arrive'' on the column ``Fundamental mathematical problems'' from the left, have no relationship with the positions of the ``connections'' which ``go out'' of that column towards the right. All connections are always ordered very logically from top to bottom.}
  \label{fig:correlation}
    \vspace{1cm}
\end{figure}

\noindent
Let us dig in the correlations between mathematical classes and application domains, via the algorithmic primitives that serve them both. Those are represented in Fig.~\ref{fig:correlation}. Notice that the results are relatively elaborate and do not immediately follow from common sense. We are going to comment on them progressively as we advance in this subsubsection. We are going to start by making a number of observations, organized per primitive, and partly based on both Figs.\ \ref{fig:matrix} and \ref{fig:correlation}, but also on general knowledge acquired during our investigation. \\

\noindent
{\bfseries$\bullet $ About the use of Grover's algorithm -- based on the primitive AA:}

\noindent
Amplitude Amplification (AA) is a major primitive, on which Grover's algorithm \cite{grover1996fast} depends {logically} (not chronologically). Based on Grover's algorithm, which uses the quantum Circuit Model of computation, many quantum search algorithms have been developed with other computational models such as Adiabatic Quantum Computation~\cite{search_adiabatic} and Quantum Walks~\mbox{\cite{search_QW1,search_qw_qa}}. The ability to search faster leads to advantages in solving problems in the ``Combinatorics''~\mbox{ \cite{aa_combi_hash,aa_combi_DS_tda, aa_combi_string,aa_combi_pathint, aa_qae_combi_counting,aa_combi_Ds_betti, aa_combi_opt_DHminim,aa_combi_datafiltering, aa_combi_search_segzdy,aa_combi_k_distinctness}}, {\mbox{``Hidden-Subgroup Problems''~\cite{aa_hsp_1,aa_hsp_2}} and ``Stochastic Processes \& statistics'' \cite{aa_sta_pagerank,aa_sta_ml_recomm,aa_sta_ml_linear_reg,aa_sta_ml_data_fitting,aa_sta_combi_kmedians_ml} classes.  \\

\noindent
{\bfseries$\bullet $ About the use of QPE -- based on the primitive QFT -- and of the QLSAs:}

\noindent
As we can see on Fig.\ \ref{fig:matrix}, QPE is one of the most popular subroutines. When aiming at solving problems in the ``Dynamical Systems'' class, QPE provides a fundamental method for estimating the ground-state energy. When aiming at solving problems in the ``Linear Algebra'' class, QPE served as a subroutine of early QLSAs (i.e., HHL). 

Further QLSAs escape QPE and make use of other primitives and subroutines instead, such as AA, QA, and QSVT \cite{Childs_qsla,aaQLSA,LinlinQSLA,costaQLSA}. Still, whatever the QLSA that we use, and since many problems can be formulated in terms of systems of linear equations, QLSAs help to solve problems in the ``Linear Algebra'', ``Optimization'', ``Stochastic Processes \& Statistics'', and ``Dynamical Systems'' classes. \\ 
%which in turn leads contributions to solving problems in the ``Optimization'' and ``Stochastic processes and statistics'' classes. The ``phase kickback'' feature of QPE provides a basic linear algebraic tool for quantum algorithms that need to manipulate the information encoded in the phase. 

\noindent
{\bfseries$\bullet $ About QAE and, in general, the mixing of QPE and AA:}

\noindent
QPE and Grover's algorithm are often used in conjuction in quantum walk-based  search algorithms for solving problems in the ``Combinatorics'' class.

QAE was also first obtained by combining QPE with  AA \cite{qae_paper}. QAE allows one to estimate the amplitude of a state within a superposition \cite{qae_paper}. It acts as a fundamental primitive for those algorithms that use amplitude encoding and require retrieving the information encoded by the amplitude. The first use that comes to mind is in counting and matching, i.e., usually in the ``Combinatorics'' class. But QAE is also used for  solving stochastic differential equations and some quantum machine-learning tasks. Moreover, further progress achieved QAE without QPE~\cite{qae_2,qae_3}. This, in turn, is used by other algorithms to perform numerical calculations and solve problems of the ``Linear Algebra'' \cite{aa_qae_opt_la2_simplex}, ``Optimization'' \cite{aa_qae_opt_la3_interior,aa_opt_qae_ml_deeplearning}, ``Stochastic Processes \& Statistics'' \cite{QAE_statistics_QMCI,QAE_statitics_classify}, and ``Dynamical Systems'' \cite{aa_qae_la_dys_NS,aa_qae_la_dys_ode, aa_qae_la1_hs,aa_la_hs_1,aa_la_hs_2} classes. \\

\noindent
{\bfseries$\bullet $ About QSVT and QSP:}

\noindent
The eventual impact of the QSVT and QSP algorithms is expected to be greater than what Fig.\ \ref{fig:matrix} indicates, not only because many of the earlier subroutine usages of HHL can be directly replaced by advanced QLSAs that use QSVT as a subroutine, but also because using QSVT for Hamiltonian Simulation (HS) could become more popular, as it can sometimes provide optimal complexity, for example for the BQP-complete problem \cite{PhysRevLett.103.150502} of Ref.\ \cite{costaQLSA}, in the LSQ sense. Moreover, powerful features such as eigenvalue transformation and filtering (of arbitrary Chebyshev polynomials) are likely to be found more uses as the exploration goes. \\

\noindent
{\bfseries$\bullet $ About HS:}

\noindent
Hamiltonian Simulation (HS), as a fundamental tool for encoding information and constructing the time evolution of the system under study, has bred a variety of approaches \cite{BE_paper, aa_la_hs_2, hs2lcu, hs3_qw, hs4, hs5_qw}. Besides the multi-purpose subroutines such as QSVT and QSP \cite{qsp_hs,GilyenUnifying}, which are not confined exclusively to HS, and excluding the Suzuki-Trotter decomposition \cite{trotter}, which we have not considered as a quantum algorithm, in Fig.\ \ref{fig:correlation} we have put all the HS algorithms under the umbrella of ``Hamiltonian Simulations'' (exactly as we did in Table \ref{tab:primitives} above, which is mentioned in the caption). Among all of these algorithms, Block Encoding emerges as the most significant primitive in our dependency network, so that we have isolated it from other HS primitives in Fig.\ \ref{fig:correlation}; we see in Fig.\ \ref{fig:correlation} that it is mostly used to solve problems in the ``Linear Algebra'' class. Generally speaking,  Fig.\ \ref{fig:correlation} shows that HS algorithms, in addition to serving problems in the ``Linear Algebra'' class, also serve problems in the ``Dynamical Systems'' class, as we may expect, most often with applications in the ``First-Principle Quantum Simulation'' domain. \\

\noindent
{\bfseries$\bullet $ About VQAs:}

\noindent
The Variational Quantum Eigensolver (VQE) \cite{vqe_review} is an algorithm designed to find the eigenvalue of an operator and it is quite often used to find the ground-state energy of a quantum system. Many quantum algorithms use the VQE as their main subroutine to solve problems in the ``Dynamical Systems'' class, with applications in chemistry or materials science, e.g., for simulating the electronic structure of a molecule \cite{Dumitrescu_2018, P_rez_Salinas_2021, 2020, Nakanishi_2019}. As a consequence, there is a large proportion of VQE uses in the ``First-Principle Quantum Simulation'' application domain, as we can see on Fig.\ \ref{fig:correlation}. However, we can also notice a good proportion of VQE uses in the ``Operational Research'' application domain, due to the growth of  variational quantum algorithms for problems in the ``Combinatorics'' and ``Optimization'' classes. It is noteworthy that following the development of the VQE, numerous Variational Quantum Algorithms (VQAs) based on the variational principle have been developed. These algorithms are capable of solving a broader range of problems in the (numerical) ``Optimization'' and ``Linear Algebra'' classes, beyond the computation of ground-state energies. \\

%\todo[inline]{write correlation analysis for VQE [Evi] and AA's[LI:done]}

%Apart from the algorithms that satisfy the definition of "primitives", some algorithms are popular subroutines that compose advanced quantum algorithms as the key components in our table, namely the QPE, QSLAs (the cluster of algorithms that solves linear system problem), QSVT, QAE, and Hamiltonian simulations (he cluster of algorithms that solves Hamiltonian simulation problem).

%In the earlier stage, QPE is frequently used for the explicit design of quantum algorithm, and lays foundation for most of the famous branch: linear algebra algorithms(HHL), combinatorial algorithms (Szegedy's QW search), statistics algorithms (QAE), Hidden subgroup algorithms (Shor), and first-principle quantum simulation (direct usage of QPE). But among all these branchs, later developed advanced algorithms try to avoid it for the sake of its cost. 
%\Cref{fig:correlation} depicts the correlation of the popularity of the primitives with the fundamental math problems and application domains. 

\noindent
{\bfseries$\bullet $ More applications of the QFT (apart from the original QPE):}

\noindent
The Quantum Fourier Transform (QFT) is one of the major primitives among quantum algorithms. With the generalization of the Fourier sampling technique \cite{bv_fs}, algorithms could rely on QFT to solve 
``Hidden-Subgroup Problems'' \cite{Imran:2022cbv,doi:10.1137/S009753970343141X,10.1145/509907.510001}, which have many applications in the Cryptanalysis domain. QFT is also used as a primitive to solve problems in the ``Dynamical Systems'' class \cite{Most_2010,Jaffali_2019}, e.g., for Hamiltonian simulation \cite{Childs:2012fnt,Childs:2010,aa_la_hs_2}. Algorithms that solve problems in the ``Linear Algebra'' \cite{PhysRevLett.103.150502,10.1145/3313276.3316366,doi:10.1137/16M1087072,Chakraborty2018ThePO,PhysRevLett.120.050502} and ``Optimization'' \cite{10.1145/3406306,10.1007/978-3-030-73879-2_22,Apeldoorn2019QuantumAF,PhysRevLett.113.130503,Wiebe2014QuantumDL,Lloyd_2014} classes also use QFT as subroutine, with applications in the ``Machine Learning \& Data Science'' and ``Classical Scientific Computing'' domains. \\

\noindent
{\bfseries$\bullet $ About QA and QUBO:}

%QFT is also used as a subroutine in association with AA in order to reach the QAE algorithm and some quantum search algorithms, which could solve statistical problems (\cite{PhysRevA.98.022321}, \cite{Paparo_2012}, \cite{Woerner_2019}, \cite{Wiebe2014QuantumAF},\cite{PhysRevLett.109.050505} ,\cite{PhysRevA.94.022342},\cite{Kerenidis2016QuantumRS}) or optimization problems (\cite{10.1145/3406306}, \cite{10.1007/978-3-030-73879-2_22},\cite{PhysRevLett.113.130503},\cite{Wiebe2014QuantumDL},\cite{Lloyd_2014}). This leads to applications in the fields of machine learning and operational research.
\noindent
Again in the context of our dependency and correlation analyses, we have put \emph{all} the Quantum Adiabatic (QA)-type algorithms under the same umbrella \cite{qa_origianl}, especially in Fig.\ \ref{fig:correlation}, and we will refer to them in the body of this paper as ``QA algorithms''. As for solving problems in the ``Linear Algebra'' and (numerical) ``Optimization'' classes, one of the crucial usages of QA algorithms is state preparation \cite{costaQLSA,qa_qlsa_2}, which is often used as a subroutine among QLSAs. Another branch of usage of QA algorithms is (heuristically) solving problems in the ``Combinatorics'' class, where the Ising-QUBO formulation \cite{ising_qubo}, as well as the the QAOA \cite{qaoa} (in the quantum Circuit Model context), give a uniform way of encoding solutions of many NP problems onto the ground state of an Ising-type spin Hamiltonian. With ingeniously designed adiabatic time scheduling, the minimum energy gap of the adiabatic Hamiltonian can be reasonably manipulated, which enables a time-complexity advantage \cite{adiabatic_review}. Notice that, due to the capacities of QA algorithms in terms of combinatorial-problem encoding and time-evolution implementation, QA algorithms give rise to some NISQ experimental implementations (i.e., Quantum Annealing, or Coherent Ising Machines)~\cite{annealing1, anealing2,annealing3, annealing4,anealing5,cim001}, which claimed to have shown evidence of quantum advantage.  \\

\noindent
{\bfseries$\bullet $ About BS:}

\noindent
Again from the perspective of NISQ devices, Boson Sampling (BS)-based algorithms~\cite{BS1,q_enable_bosonsamp}, including Gaussian BS-based algorithms, also claimed to have shown evidence of quantum advantage in solving problems in the ``Combinatorics'' and ``Stochastic Processes \& Statistics'' classes \cite{GBS_prl,GBS1,gbs2,gbs2.5,gbs3,gbs4,gbs5}. \\

\noindent
{\bfseries$\bullet $ About the variety of quantum test algorithms:}

\noindent
Finally, quantum test algorithms form an indispensable toolbox for quantum-algorithm research. The Swap Test is used to compare two states and can be used for fidelity checks \cite{q_enable_swap_test}. This solves problems in the ``Linear Algebra'' class. The Hadamard Test is used to sample the expectation of acting with a given operator on some input state, which enables to tackle problems in the ``Stochastic Processes \& Statistics'' class \cite{q_enable_hadamard_test}. In addition, the Hadamard Test is used to calculate the distance between two operators, which is meaningful for evaluating some quantum algorithms that update gate parameters iteratively and solve problems in the ``Linear Algebra'' class. \\

\noindent
{\bfseries$\bullet $ Discussing Fig.\ \ref{fig:correlation}:}

\noindent
Overall, it is noteworthy that the quantum subroutines do not entirely dictate the relationship between application domains and mathematical classes. Instead, formulating application-domain problems into quantum-solvable mathematical problems by modeling and reduction is a crucial part of exploring practical applications of quantum computing. Moreover, a given, general field of application, actually often spans across several application domains: for instance, the task of drug design spans across application domains such as ``Operational Research'', ``First-Principle Quantum Simulation'', and ``Machine Learning and Data Science''; indeed and \mbox{hence~--~since} we now merge (i) this large spectrum of applications with (ii) its influence on the precise mathematical problems to be solved, just to give an example of the fact that a strict distinction between both is not always that relevant --, depending on the scale the chemical properties of a drug can be modeled and analyzed either from first principles, semi-empirically, or entirely numerically, and the pharmacological and toxicological properties of the drug could also be optimized along these paths \cite{drug_potential}. A counter example example is the Cryptanalysis application domain. The relevance of this application domain to mathematical problems is quite singular: quantum algorithms for solving hidden-subgroup problems are almost exclusively applied in this field, and they are based on QFT. All that being said, despite the complexity of Fig.\ \ref{fig:correlation}, when faced with computationally intensive tasks, our typology could provide an effective perspective, offering potential quantum-computing users a path to find quantum speed-up solutions.

\subsection{NISQ and secondary criteria} \label{subsec:NISQ}

\noindent
{\bfseries$\bullet $ NISQ/LSQ criterion with respect to the two main criteria}

\noindent
Our analysis of the ``NISQ/LSQ'' criterion resulted in two figures illustrating the percentage of NISQ/LSQ algorithms categorized by mathematical classes (Fig.\ \ref{fig:math_nisq_final_paper}) or application domains~(Fig.~\ref{fig:Nisq_application_final}).
Judging by Fig.\ \ref{fig:Nisq_application_final}, it seems that the most promising application domain is ``First-Principle Quantum Simulation'', likely due to the variational algorithms that have been developed lately for this purpose, although full quantum simulation will still require LSQ devices. The ``Classical Scientific Computing'' application domain also seems promising, which is primarily due to the development of variational linear-algebra algorithms (also called QLSAs), as well as of other algorithms tackling dynamical-system problems. In contrast, there is a limited number of NISQ algorithms in the ``Cryptanalysis'' application domain, which is due to the fact that this domain prominently depends on the QFT.
The large percentage of ``Stochastic Processes \& Statistics'' algorithms falling into the NISQ category is due to the probabilistic nature of stochastic process problems, which are often tackled with sampling algorithms and seem more compatible with noisy devices, showing great promise in achieving quantum-computational advantages. \\ \\

\begin{figure}[h!] 
  \centering       
  %\hspace{-2cm}
  \includegraphics[width=1\textwidth]{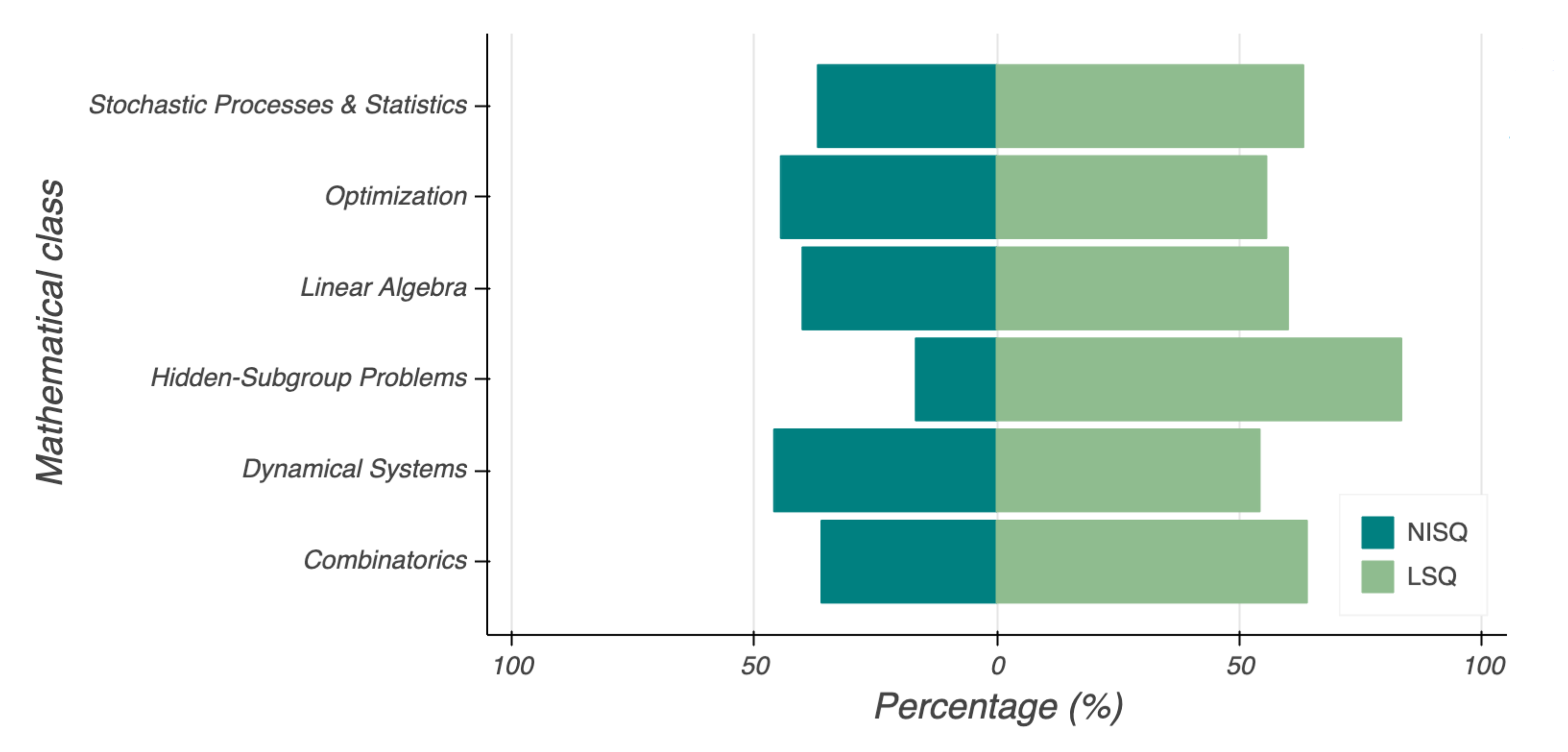} 
  \caption{{\bfseries NISQs per Mathematical Class.} Percentage of NISQ/LSQ algorithms per mathematical class. 
  \label{fig:math_nisq_final_paper}
  }
  \vspace{1cm}
\end{figure}

\begin{figure}[h!] 
  \centering       
  %\hspace{-2cm}
  \includegraphics[width=1\textwidth]{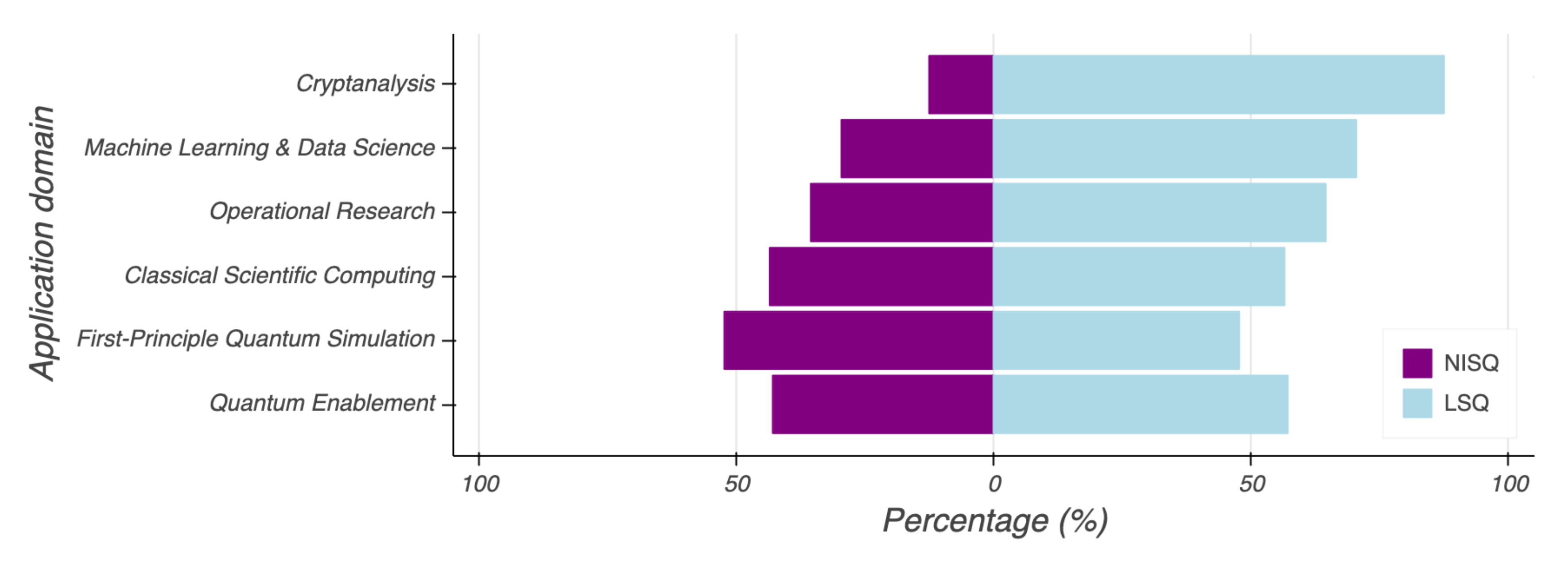} 
  \caption{{\bfseries NISQs per Application Domain.} Percentage of NISQ/LSQ algorithms per application domain.}
  \label{fig:Nisq_application_final}
  \vspace{1cm}
\end{figure}

%\vspace{2cm}

\noindent
{\bfseries$\bullet $ Other criteria}

\noindent
Let us close this analysis section by commenting about some correlations unraveled by the classification table, and involving secondary criteria such a being ``Heuristic/Proven'', ``Oracular'', ``Promise'', or the subject of ``Partial dequantization''. 

We notice that NISQ is correlated with being a -heuristic algorithm. This is as one would expect. The general philosophy behind NISQ algorithm design is to obtain results in spite of errors and limited resources --  which, in particular, unfortunately implies that any advantage is going to be hard to prove in this context. Conversely, LSQ is correlated with Proven algorithms, showing that whenever researchers are working in this more idealized setting, they do strive for proofs of quantum advantage. Similarly, the correlation between being Oracular or Promise, which plant an even more idealized setting, with being Proven, comes to reinforce this understanding. Let us close this loop by noticing that there are almost no Oracular NISQ algorithms -- again, because the NISQ rationale to come up with things that work right now on present-day hardware, is somewhat orthogonal to the mathematically inclined considerations of there existing some abstract oracle.

Within LSQ, the mathematical class ``Combinatorics'' is correlated with the algorithm being Oracular. This is likely explained by the fact that the AA algorithm (and search algorithms in general) are Oracular. The mathematical class ``Dynamical Systems'', on the other hand, is anti-correlated with being Oracular, likely reflecting the importance of the QPE and Non-oracular VQE  algorithms in this field. 

Partial dequantization mainly affects Proven LSQ algorithms, and mainly from the mathematical class ``Linear Algebra'' but not only (some ``Optimization'', ``Combinatorics'' and ``Dynamical Systems'' problems are impacted), and mainly from the ``Machine Learning \& Data Science'' application domain but not only (some ``Operational Research'' and ``First-Principle Quantum Simulation'' applications are affected). This is mostly explained by the underlying use of the QPE and HHL algorithms. 

The ``Sampling'' criterion does not show any significant correlation.

\label{sec:sec2}
%\chapter{Conclusion}
\section{Conclusion}
\label{sec:conclusion}
\noindent
Quantum computing, driven by its exploitation of quantum-mechanical phenomena, stands poised to revolutionize computation. This paper has undertaken the task of mapping the ever-expanding landscape of quantum algorithms. To handle this rapidly moving field, we built a large sample of significant quantum algorithms based on a literature-review approach, seeking to capture both the most established algorithms and the most recent trends. 

We analyzed the dependencies of these algorithms, drawing up their family tree and thereby making it apparent which are the true primitives and the most popular subroutines that are powering up quantum advantage. 

For each of these algorithms, we listed the ``Fundamental mathematical problem'' addressed, from which we then identified 6 larger classes. The same bottom-up approach was followed for the real-world ``Applications'' of these algorithms, which we first listed and then gathered into 6 application domains. Analyzing algorithmic trends over time, we observed shifts in the development of quantum algorithms, with certain mathematical classes and application domains gaining prominence. Notably, the domains of ``Machine Learning \& Data Science'', of ``First-Principle Quantum Simulation'', and of ``Operational Research'', demonstrated substantial growth.

We took a closer look at the significance of the major primitives and/or subroutines in this unravelling story: Amplitude Amplification (AA), Quantum Phase Estimation (QPE), Quantum Linear-System Algorithms (QLSAs), Quantum Singular-Value Transformation (QSVT), Quantum Signal Processing (QSP), Hamiltonian Simulation (HS), Variational Quantum Eigensolver~(VQE), Quantum Fourier Transform (QFT), Quantum Adiabatic (QA) algorithms, Ising-QUBO formulation, Boson Sampling (BS), Testing. 

Other criteria (``Sampling'', ``Promise'', ``Heuristic/Proven'', ``NISQ/LSQ'') were gathered and their correlations analyzed. We hope that all of these insights will provide guidance for students, researchers, companies and decision makers, navigating the NISQ era. For instance, the ``First-Principle Quantum Simulation'', ``Classical Scientific Computing'', and
``Stochastic Processes \& Statistics'' application domains where identified as particularly promising in the short term.

Whilst we did our best to offer a good snapshot and analysis of the current landscape, this work will no doubt require periodic updates in order to accommodate continuing developments: our data and codes are made publicly available for anyone willing to undertake this task. During our research, we became aware of a similar effort presented as an arXiv preprint \cite{dalzell2023quantum}, which is more descriptive and topic-focused in nature, while try to focus on the bigger picture. Independent perspectives and methods are needed to illuminate such a vast and complex landscape, and constitute an essential ingredient of the scientific method. Future research could try and incorporate algorithmic complexity and performance bounds as a further criterion (of some classification table similar to ours, for example).

%\todo[inline]{cite the paper AMAZON'S paper, at the end of the project we realized about a similar ongoing project}

%\addcontentsline{toc}{chapter}{References}
%\section{References}
\newpage
%\printbibliography[title={References}]
%\nocite{apsrev41Control}
%\bibliographystyle{apsrev4-1}
%\bibliographystyle{unsrtnat}
%\bibliography{ref.bib} 

\appendix
\section{Classification table} \label{app:classification_table}

\noindent
All the columns of the classification table we provide in this appendix have been explained in Sec.~\ref{sec:Methodology} -- there are, apart from that which gives the name of the algorithm, 11 of them, plus a column providing the link towards the paper that dequantizes the algorithm whenever applicable~--, and used in our analysis of \mbox{Sec.\ \ref{sec:analysis}}, except for three extra columns, which have not been used in our analysis, and provide, respectively:
\begin{enumerate}
    \item The link towards the arXiv version of the paper whenever available ;
    \item The link towards the peer-reviewed version of the paper whenever available ;
    \item The date of publication, which is the minimum date between the arXiv and the peer-reviewed versions of the paper.
\end{enumerate}

\newpage
\begin{figure}[ht]
  \centering
  \includepdf[pages=1, linktodoc=true, width=\textheight, angle=270]{Final_table.pdf}
  %\caption{Classification table}
  \label{fig:Classification table}
\end{figure}

\newpage
\begin{figure}[ht]
  \centering
  \includepdf[pages=2, linktodoc=true, width=\textheight, angle=270]{Final_table.pdf}
  %\caption{Classification table}
  %\label{fig:Classification table}
\end{figure}
\newpage
\begin{figure}[ht]
  \centering
  \includepdf[pages=3, linktodoc=true, width=\textheight, angle=270]{Final_table.pdf}%  %\caption{Classification table}%  \label{fig:Classification table}%\end{figure}
  \end{figure}
\newpage
\begin{figure}[ht]
  \centering
  \includepdf[pages=4, linktodoc=true, width=\textheight, angle=270]{Final_table.pdf}
  %\caption{Classification table}
  %\label{fig:Classification table}
\end{figure}

\newpage
\begin{figure}[ht]
  \centering
  \includepdf[pages=5, linktodoc=true, width=\textheight, angle=270]{Final_table.pdf}
 % \caption{Classification table}
  %\label{fig:Classification table}
\end{figure}

\newpage
\begin{figure}[ht]
  \centering
  \includepdf[pages=6, linktodoc=true, width=\textheight, angle=270]{Final_table.pdf}
 % \caption{Classification table}
  %\label{fig:Classification table}
\end{figure}

\newpage
\begin{figure}[ht]
  \centering
  \includepdf[pages=7, linktodoc=true, width=\textheight, angle=270]{Final_table.pdf}
 % \caption{Classification table}
  %\label{fig:Classification table}
\end{figure}

\newpage
\begin{figure}[ht]
  \centering
  \includepdf[pages=8, linktodoc=true, width=\textheight, angle=270]{Final_table.pdf}
 % \caption{Classification table}
  %\label{fig:Classification table}
\end{figure}

\newpage
\begin{figure}[ht]
  \centering
  

\section*{Supplementary material}

| Quantum algorithm | Main subroutine | Fundamental mathematical problem | Applications | Heuristic/ Proven | NISQ/ LSQ | Oracular | Promise | Sampling | Partial dequantization | Dequan tization link | Computational model | Google-Scholar citations | Link to arXiv paper | Link to peer-reviewed paper | Date |
|---|---|---|---|---|---|---|---|---|---|---|---|---|---|---|---|
| Q. k-means Algo. | Swap Test, Dürr & Høyer's Minimization Algo., Adiabatic PageRank Algo. | Stochastic Processes & Statistics: Classification of data (clustering) | Machine Learning & Data Science: Unsupervised learning, Data analysis | Proven | LSQ | Yes | No | No | No | | Adiabatic Quantum Computation | 678 | https://arxiv.org/abs/13 07.0411 | None | 04/11/2013 |
| Q. k-medians Algo. | Swap Test, Dürr & Høyer's Minimization Algo. | Stochastic Processes & Statistics: Classification of data (clustering) | Machine Learning & Data Science: Unsupervised learning, Data analysis | Proven | LSQ | Yes | No | No | Yes | | Circuit Model | 102 | None | https://dl.acm.org/doi/10 | 20/06/2007 |
| Q. Nearest-Neighbor Classification | Grover, QAE | Stochastic Processes & Statistics: Binary classification (clustering), Calculating Euclidean distance between vectors or calculation of inner product | Machine Learning & Data Science: Handwriting recognition | Proven | LSQ | Yes | No | No | No | | Circuit Model | 224 | https://arxiv.org/pdf/14 01.2142.pdf | 5555/287 2871400 | 18/07/2014 |
| Q. Hierarchical Clustering Algo. | Primitive | Stochastic Processes & Statistics: Classification of data (divise clustering approach) | Machine Learning & Data Science: Unsupervised learning | Proven | NISQ | No | No | No | No | | Circuit Model | 1407 | https://arxiv.org/pdf/18 05672. | Yes, requires training data not to be atypical of Haar-random unit vectors | 10/07/2014 |
| Q. Support-Vector Machine (QSVM) | HHL, Swap Test | Optimization, Linear Algebra: Linear discrimination problems (which is subcategory of pattern/linear classification) | Machine Learning & Data Science: Supervised learning, Pattern Recognition | Proven | LSQ | Yes | Yes, low-rank matrices | No | Yes | https://doi.org/10.1145/3357713, 3384314, | Circuit Model | 1407 | https://arxiv.org/abs/13 07.0471 | ad/pdf/78 056672. | 10/07/2014 |
| Q. Variational Support-Vector Machine | Inspired by VQE | Optimization, Linear Algebra: Linear discrimination problems (which is subcategory of pattern/linear classification) | Machine Learning & Data Science: Supervised learning, Pattern recognition | Heuristic | NISQ | No | No | No | No | | Circuit Model | 1214 | https://arxiv.org/pdf/18 04.11326. | nature.com/articl es/s4158 | 05/06/2018 |
| Q. Data Fitting | HHL, AA, Swap Test | Stochastic Processes & Statistics: Fitting data to theoretical models, Determination of the quality of a least-squares fit over a data set; Least-squares fitting - Estimating Fit Quality | Machine Learning & Data Science: Classification, Inference, Training deep restricted & full Boltzmann machine | Proven | LSQ | Yes | Yes | Algo. 1: No Algo. 2: Yes | No | | Circuit Model | 496 | https://arxiv.org/pdf/12 04.5242 | aps.org/pdf/abstract/10.1103/Phy | 03/07/2012 |
| Q. Distance Classifier (QDC) | Primitive | Optimization: Computation of distance between datapoints, Classification | Machine Learning & Data Science: Supervised learning, Pattern recognition | Proven | NISQ | No | Yes | No | No | | Circuit Model | 219 | https://arxiv.org/article | https://iopscience.iop. | 28/08/2017 |
| Q. Principal-Component Analysis | HHL, Swap Test | Optimization, Linear Algebra: Dimensionality reduction | Machine Learning & Data Science: Unsupervised machine learning, Data analysis | Proven | LSQ | No | Yes | Yes | Yes | https://doi.org/10.1103/Phys RevLett | Quantum Random-Access Memory | 1098 | https://arxiv.org/abs/13 07.0401 | nature.com/article | 16/09/2013 |
| Q. Linear Regression | HHL, AA | Optimization, Linear Algebra: Linear Regression using least-squares optimization | Machine Learning & Data Science: Supervised learning, Pattern recognition, Prediction | Proven | LSQ | No | Yes | No | Yes | https://doi.org/10.1145/3357713 | Circuit Model | 234 | https://arxiv.org/pdf/16 01.07823. | aps.org/pdf/abstract/ | 04/08/2016 |
| Q. Pattern Maching | Dihedral HSP Algo. | Stochastic Processes & Statistics: Pattern matching | Machine learning & Data science, Classical Scientific Computing: Text and image processing, bioinformatics | Proven | LSQ | Yes | No | No | No | | Circuit Model | 64 | https://arxiv.org/pdf/14 08.1816.pdf | springer.com/article/10 | 26/08/2015 |
| Q. Algo. for Recommendation Systems | QPE, AA | Stochastic Processes & Statistics: Matrix Sampling, Linear Algebra: Singular-value decomposition | Machine Learning & Data Science: Personalized recommendations to individual users | Proven | LSQ | No | No | Yes | Yes | http://dx.doi.org/10.1145/3313 | Circuit Model | 344 | https://arxiv.org/pdf/16 09.13581.pdf | https://drops.dagstuhl. | 22/09/2017 |
| Q. Algo. for Topological and Geometric Analysis of Data (LGZ) | QPE, Grover | Combinatorics: Estimate Betti numbers in persistent homology and finding eigen vectors and eigenvalues of combinatorial Laplacian | Machine Learning & Data Science: Topological data analysis (information retrieval) | Proven | NISQ | No | No | No | Yes | | Circuit Model | 208 | https://arxiv.org/pdf/14 08.3106.pdf | www.nature.com/articl es/ncomm | 13/08/2014 |
| Q. Algo. for Persistent Betti Numbers | AA, QSVT | Combinatorics: Algebraic topology (estimation of persistent Betti numbers) | Machine Learning & Data Science | Proven | LSQ | Yes | No | No | Yes | | Machine Learning & Data Science | 14 | https://arxiv.org/pdf/220 4.10592 | https://quantum-journal | 06/12/2022 |

| Quantum algorithm | Main subroutine | Fundamental mathematical problem | Applications | Heuristic/ Proven | NISQ/ LSQ | Oracular | Promise | Sampling | Partial dequantization | Dequantization link | Computational model | Google-Scholar citations | Link to arXiv paper | Link to peer-reviewed paper | Date |
|---|---|---|---|---|---|---|---|---|---|---|---|---|---|---|---|
| Black-Box HS and Unitary Implementation | QPE, AA | Dynamical Systems: Hamiltonian simulation. Linear Algebra: Implementation of unitary transformation | First-Principle Quantum Simulation | Proven | LSQ | Yes | No | No | No | | Quantum Walks | 212 | https://arxiv.org/abs/0910.4157 | None | 02/07/2011 |
| Sparse HS | Primitive | Dynamical Systems: Hamiltonian simulation | First-Principle Quantum Simulation | Proven | LSQ | Yes | No | No | No | | Circuit Model | 723 | https://arxiv.org/abs/... | None | 07/02/2006 |
| QW-HS by Childs | Primitive | Dynamical Systems: Hamiltonian simulation (sparse time-independent & non-sparse) | First-Principle Quantum Simulation | Proven | LSQ | Yes | No | No | No | | Quantum Walks | 416 | https://arxiv.org/abs/08... | https://doi... | 10/10/2009 |
| QSP | QPE | Dynamical Systems: Hamiltonian simulation (sparse time-independent) | First-Principle Quantum Simulation | Proven | LSQ | Yes | Yes | No | No | | Quantum Walks | 538 | https://arxiv.org/abs/16... | https://doi.org/10.1145/3357... | 20/12/2016 |
| LCU | Primitive | Dynamical Systems: Hamiltonian simulation (time-independent sparse Hamiltonian simulation) | First-Principle Quantum Simulation | Proven | LSQ | No | Yes | No | Primitive is | | Circuit Model | 396 | https://arxiv.org/abs/12... | None | 27/02/2012 |
| QW-LCU-HS | QW-HS by Childs, LCU, AA | Dynamical Systems: Hamiltonian simulation | First-Principle Quantum Simulation | Proven | LSQ | Yes | Yes | No | No | https://doi.org/10.1145/3357... | Quantum Walks | 354 | https://arxiv.org/abs/15... | https://doi... | 07/12/2015 |
| Ising-QUBO | QA | Combinatorics, Optimization: Quadratic unconstraint binary optimization | Operational Research: Classical Scientific Computing | Heuristic | NISQ | No | No | No | Yes | Local search | Adiabatic Quantum Computation | 1884 | https://arxiv.org/abs/13... | https://www.frontiersin... | 23/02/2013 |
| Adiabatic Search | QA, Inspired by Grover | Combinatorics: Binary classification on finite sets | Machine Learning & Data Science: Data base search, Operational Research | Proven | LSQ | Yes | No | No | No | | Adiabatic Quantum Computation | 718 | https://arxiv.org/qu... | https://doi... | 03/07/2001 |
| QA-QW-Hybrid Noise-Robust Search | Adiabatic Search, SKW's QW Search | Combinatorics: Binary classification on finite sets | Machine Learning & Data Science: Data base search, Operational Research | Proven | NISQ | Yes | No | No | No | | Adiabatic Quantum Computation, Quantum Walks | 38 | https://arxiv.org/abs/17... | https://doi... | 01/09/2017 |
| Adiabatic PageRank Algo. | QA | Stochastic Processes & Statistics: Measure the relative importance of an element within a set | Operational Research: Search engine optimization, Machine Learning & Data Science: Data mining, info retrieving | Proven | LSQ | No | Yes, advantage only for top rank log(n) pages | Yes | No | | Adiabatic Quantum Computation | 104 | https://arxiv.org/abs/11... | https://link.aps.org/doi/10... | 29/09/2011 |
| Q.-Annealing-Based Graphical Game Pure Nash Equilibrium Algo. | Ising-QUBO | Combinatorics, Optimization: Nash equilibrium | Operational Research: Strategic decision making | Heuristic | NISQ | No | No | No | No | | Adiabatic Quantum Computation, Circuit Model | 2 | 1903.064 541A Quantum... | None | 15/03/2019 |
| Q. Adiabatic Factorization | Inspired by QA, Inspired by Shor | Hidden-Subgroup Problems: Integer factorization. Dynamical Systems: Lowest energy state of nuclear spins | Cryptanalysis | Proven | LSQ | No | No | No | No | | Adiabatic Quantum Computation, Circuit Model | 185 | https://arxiv.org/abs/08... | https://doi... | 14/08/2008 |
| Q.-Annealing Factorization | Ising-QUBO | Hidden-Subgroup Problems: Integer factorization | Cryptanalysis | Heuristic | NISQ | No | No | No | No | | Adiabatic Quantum Computation | 91 | https://arxiv.org/abs/16... | https://doi.org/10.1038/sre... | 16/06/2016 |
| Q. Adiabatic Large-Scale Binary Classifier Training | Ising-QUBO | Combinatorics, Optimization: Binary classification on finite sets | Machine Learning & Data Science | Heuristic | LSQ | No | No | No | No | | Adiabatic Quantum Computation | 88 | https://arxiv.org/abs/09... | None | 04/12/2009 |
| Q. Annealing for Multijet Clustering | Ising-QUBO | Stochastic Processes & Statistics: Clustering | Classical Scientific Computing: Multijet clustering in high energy physics | Heuristic | NISQ | No | No | No | No | | Adiabatic Quantum Computation, Circuit Model | 17 | https://arxiv.org/abs/20... | https://www.sciencedi... | 28/12/2020 |
| Adiabatic QLSA | QAOA, QSVT | Linear Algebra: Linear system of equations | Machine Learning & Data Science, Classical Scientific Computing: Computational Mechanics | Proven | NISQ | No | Yes, same as HHL | No | No | | Adiabatic Quantum Computation, Circuit Model | 71 | https://arxiv.org/pdf/ful... | https://doi.org/10... | 04/03/2022 |
| QAOA | QA | Combinatorics: Max-cut, Search subset of the n-bit strings | Operational Research | Heuristic | NISQ | No | No | No | No | | Adiabatic Quantum Computation, Circuit Model | 2433 | https://arxiv.org/abs/14... | None | 14/11/2014 |
| Reverse-Q.-Annealing Approach to Portfolio Optimization Problems | Ising-QUBO | Combinatorics, Optimization: Quadratic unconstrained optimization problem | Operational Research: Finance: Portfolio optimization | Heuristic | NISQ | No | No | No | No | | Adiabatic Quantum Computation | 163 | https://arxiv.org/pdf/18... | https://link.springer.com/articl... | 19/10/2018 |

| Quantum algorithm | Main subroutine | Fundamental mathematical problem | Applications | Heuristic/ Proven | NISQ/ LSQ | Oracular | Promise | Sampling | Partial dequantization | Dequantization link | Computational model | Google-Scholar citations | Link to arXiv paper | Link to peer-reviewed paper | Date |
|---|---|---|---|---|---|---|---|---|---|---|---|---|---|---|---|
| Coherent-Ising-Machine Max-Cut Heuristic | Ising-QUBO | Combinatorics: Max-cut | Machine Learning & Data Science: Binary classification. Operational Research. | Heuristic | NISQ | No | No | No | No | | Adiabatic Quantum Computation | 312 | https://arxiv.org/abs/13 11.2956 | https://journals.aps.org/pra/a | 26/08/2013 |
| Quantum Ising Shortest-Vector Problem Algo. | Ising-QUBO | Hidden-Subgroup Problems, Combinatorics, Optimization: Shortest Vector Problem | Cryptanalysis. | Heuristic | NISQ | No | No | No | No | | Adiabatic Quantum Computation | 22 | https://arxiv.org/abs/20 | | 24/06/2020 |
| Q.-Enhanced Markov-Chain Monte Carlo | Primitive | Stochastic Processes & Statistics: Boltzmann sampling | Machine Learning & Data Science. | Heuristic | NISQ | No | Yes | No | No | | Circuit Model | 14 | https://arxiv.org | https://www.nature | 23/03/2022 |
| Q. Monte-Carlo Integral | QAE | Stochastic Processes & Statistics: Approximate numerical integration | Machine Learning & Data Science. Classical Scientific Computing. Operational Research | Proven | LSQ | No | No | No | | | Circuit Model | 337 | https://arxiv.org/abs/15 | https://royalsociety publishing | 08/09/2015 |
| Durr & Høyer's Minimization Algo. | Grover | Optimization: Searching the index with the minimum value | Machine Learning & Data Science | Proven | LSQ | No | No | No | | | Circuit Model | 517 | None | | 18/07/1996 |
| (Approximate) Q. Counting | QAE | Combinatorics: Counting the number of elements that fulfill some specific requirements | Machine Learning & Data Science: Computer vision - histogram-based image processing | Proven | LSQ | No | No | No | | | Circuit Model | 553 | https://arxiv.org/abs/qu ant-ph/98050 | https://link.springer.com/chap ter/10 | 27/05/1998 |
| SKW's QW Search | Variant of Grover | Combinatorics: Search elements on a given graph | Machine Learning & Data Science: Data base search. Operational Research | Proven | LSQ | Yes | Yes, only on hypercube and 2-d lattice | No | | | Quantum Walks | 1292 | https://arxiv.org/abs/qu | https://journals.aps. | 9/10/2002 |
| Szegedy's QW | Variant of Grover | Combinatorics: Searching elements in a set | Machine Learning & Data Science: Data base search. Operational Research | Proven | LSQ | Yes | No | No | | | Quantum Walks | 712 | None | https://ieeexplore.ieee. | 17/10/2004 |
| MNRS's QW Search | Szegedy's QW, QPE | Combinatorics: Searching elements on a graph, more general than Tulsi's | Machine Learning & Data Science: Data base search. Operational Research | Proven | LSQ | Yes | No | No | | | Quantum Walks | 463 | https://arxiv.org/abs/qu ant | https://dl.acm.org/doi/ab | 01/06/2007 |
| Tulsi's QW Search on a 2D Grid | Inspired by Szegedy's QW | Combinatorics: Searching elements on a 2D grid in O(sqrt(N) sqrt(log N)) rather than O(sqrt(N) log N) as the previous algorithms (such as Szegedy's) | Machine Learning & Data Science: Data base search. Operational Research | Proven | LSQ | Yes | No | No | | | Quantum Walks | 219 | https://arxiv.org/abs/08 01.0497 | https://journals.aps.org/pra/a bstract/10 | 09/07/2008 |
| Q. Algo. for String Matching | Durr & Høyer's Minimization Algo. | Combinatorics: String matching | Machine Learning & Data Science: Pattern matching, recommendation system. Classical Scientific Computing. Cosmology data processing | Proven | LSQ | No | No | No | | | Quantum Walks | 80 | https://arxiv.org/abs/qu ant-ph/00110 | https://www.sciencedir ect.com/scien | 13/11/2000 |
| Q. PageRank (QW&Grover based) | Variant of Grover | Stochastic Processes & Statistics: Measure the relative importance of an element within a set | Operational Research: Search engine optimization. Machine Learning & Data Science: Data mining, information retrieving | Heuristic | LSQ | No | No | No | | | Quantum Walks | 141 | https://arxiv.org/abs/qu ant | https://www.nature.com/articl es/srep00 | 09/12/2011 |
| Speedup in Mixing | Szegedy's QW, AA | Stochastic Processes & Statistics: Sampling | Operational Research | Proven | LSQ | Yes | No | Yes | No | | Quantum Walks | 116 | https://arxiv.org/abs/08 | https://journals.aps. | 31/10/2008 |
| QAE | QPE, AA | Linear Algebra: Estimate the amplitude of a substate of a superpositioned state | Quantum Enablement | Proven | LSQ | Yes | No | No | No | | Circuit Model | 1690 | https://arxiv.org/abs/qu ant | https://www.ams.org/books | 15/05/2000 |
| Iterative QAE | AA | Linear Algebra: Estimate the amplitude of a substate of a superpositioned state | Quantum Enablement | Proven | LSQ | Yes | No | No | No | | Circuit Model | 111 | https://arxiv.org/pdf/19 | | 11/12/2019 |
| Maximum Likelihood QAE | AA | Linear Algebra: Estimate the amplitude of a substate of a superpositioned state | Quantum Enablement | Proven | NISQ | Yes | No | No | No | | Circuit Model | 196 | https://arxiv.org/abs/19 02.10246 | https://link.springer.com/article | 23/04/2019 |
| Boolean Circuit Evaluation | QPE | Combinatorics: Circuit evaluation | Cryptanalysis/Cryptography. Machine Learning & Data Science. Operational Research: Circuit optimization | Proven | LSQ | Yes | Yes | No | No | | Quantum Walks | 231 | None | https://doi.org/10. 1137/080 | 27/11/2007 |

| Quantum algorithm | Main subroutine | Fundamental mathematical problem | Applications | Heuristic/ Proven | NISQ/ LSQ | Oracular | Promise | Sampling | Partial dequantization | Dequantization link | Computational model | Google-Scholar citations | Link to arXiv paper | Link to peer-reviewed paper | Date |
|---|---|---|---|---|---|---|---|---|---|---|---|---|---|---|---|
| St-connectivity in graphs | AA, QPE | Combinatorics: Graph connectivity (cycle detection / bipartiteness testing) | Operational Research, Network analysis, graph coloring, optimization. Classical Scientific Computing: Capacitance of a circuit. | Proven | LSQ | Yes | Yes | No | No | | Quantum Walks | 26 | https://arxiv.org/abs/1610.00581 | https://www.mitpressjournals.com/doi/abs/10.1... | 03/10/2016 |
| k-distinctness | Variant of Grover | Combinatorics: Element distinctness | Cryptanalysis, Machine Learning & Data Science; Data-base search. Operational Research | Proven | LSQ | No | No | No | No | | Quantum Walks | 1052 | https://arxiv.org/abs/qu... | https://epubs.siam... | 01/11/2003 |
| Hilbert-Schmidt Test | Primitive | Linear Algebra: Complex Unitary distance evaluation | Quantum Enablement | Proven | NISQ | No | No | No | No | | Circuit Model | 311 | https://arxiv.org/abs/18... | https://dl.acm.org/doi/ab... | 02/07/2018 |
| Hadamard Test | Primitive | Linear Algebra, Stochastic Processes & Statistics: Calculate the expectation value of a unitary projecting on a state. | Quantum Enablement | Proven | NISQ | No | No | No | No | | Quantum Walks | 359 | https://arxiv.org/abs/quant-ph/0511109 | https://doi.org/10.1145/113... | 09/11/2005 |
| Swap Test | Primitive | Linear Algebra: Compare two states | Quantum Enablement | Proven | NISQ | No | No | No | No | | Circuit Model | 239 | https://arxiv.org/pdf/20.10488 | https://epubs.org/pres... archi... | 25/04/1996 |
| Variational Q. Fisher-Information Estimation | VQE | Dynamical Systems: Estimating quantum Cramér-Rao bound. Linear Algebra: Comparing two open-system quantum operation | First-Principle Quantum Simulation: Simulating quantum sensors and evaluating its reliability | Heuristic | NISQ | No | No | No | No | | Circuit Model | 54 | https://arxiv.org/pdf/... | https://quantum-journal... | 20/10/2020 |
| Variational Q. Fidelity Estimation | Variational Q. State Diagonalization | Linear Algebra: Estimating fidelity of two states | Quantum Enablement | Heuristic | NISQ | No | No | No | No | | Circuit Model | 115 | https://arxiv.org/pdf/19... ne.09053 | https://quantum-journal... | 21/06/2019 |
| BS | Primitive | Stochastic Processes & Statistics: Boson sampling | First-Principle Quantum Simulation, Quantum Enablement | Proven | NISQ | No | No | Yes | No | | Linear Optical Quantum Computing | 1485 | https://arxiv.org/... | https://dl.acm... | 14/11/2010 |
| Gaussian BS (GBS) | Variant of BS | Stochastic Processes & Statistics: Boson sampling | First-Principle Quantum Simulation | Proven | NISQ | No | No | Yes | No | | Linear Optical Quantum Computing | 326 | https://arxiv.org/abs/13... | https://journals... | 19/05/2013 |
| GBS Perfect Matching | Gaussian BS (GBS) | Combinatorics: Perfect matching | Classical Scientific Computing, Operational Research | Proven | NISQ | No | No | Yes | No | | Linear Optical Quantum Computing | 95 | https://arxiv.org/abs/17... | https://journals... | 19/12/2017 |
| GBS Dense k-subgraph | GBS Perfect Matching | Combinatorics: Dense k-subgraph identification | Operational Research, Machine Learning & Data Science | Heuristic | NISQ | No | No | Yes | No | | Linear Optical Quantum Computing | 110 | https://arxiv.org/pdf/18... ans... | https://journals... | 28/03/2018 |
| Weighted Maximum / Maximum Clique via GBS | GBS Dense k-subgraph | Combinatorics, Optimization: Weighted maximum / maximum clique | Classical Scientific Computing: Molecular docking, RNA folding prediction | Heuristic | NISQ | No | No | Yes | No | | Linear Optical Quantum Computing | 99 | https://arxiv.org/abs/19... | https://www.science... | 01/02/2019 |
| Graph Isomorphism/Similarity Evaluation via GBS | Gaussian BS (GBS) | Combinatorics: Graph isomorphism/similarity | Classical Scientific Computing: Chemical identification. Machine Learning & Data Science: Data mining, Image recognition. Operational Research | Proven | NISQ | No | No | Yes | No | | Linear Optical Quantum Computing | 45 | https://arxiv.org/abs/1810.10...441 Graph isomorphism_and_Gaussian... | https://www.degruyter.com/document/doi/10... | 24/10/2018 |
| GBS Matrix Point Process | Gaussian BS (GBS) | Stochastic Processes & Statistics: Matrix point process | First-Principle Quantum Simulation, Machine Learning & Data Science: Information retrieval | Proven | NISQ | No | No | Yes | No | | Linear Optical Quantum Computing | 34 | https://arxiv.org/pdf/19... | https://journals... | 27/06/2019 |
| BS for Vibronic Spectra (Absorption Spectra) | BS | Dynamical Systems: Computing Franck Condon profile of a molecule system (vibronic spectra) | Computing molecular vibronic spectra, including complicated effects such as Duschinsky rotations | Heuristic | NISQ | No | No | Yes | No | | Linear Optical Quantum Computing | 319 | https://arxiv.org/abs/1412.8427 | https://pubs.acs.org/doi/abs/10... | 29/12/2014 |
| Max Hafnian via GBS | Gaussian BS (GBS) | Combinatorics, Optimization: Enumerating max Hafnian of given input | Operational Research: Graph optimization | Heuristic | NISQ | No | No | Yes | No | | Linear Optical Quantum Computing | 47 | https://arxiv.org/abs/18... | https://journals.aps... | 28/03/2018 |
| Optimal 2D Fermionic QFT | Primitive | Dynamical Systems: Fourier transform | First-Principle Quantum Simulation | Proven | NISQ | No | No | No | No | | Circuit Model | 162 | https://arxiv.org/abs/1711.053 95... | https://doi.org/10.1103/PhysRevAppli... | 15/11/2017 |
| Strongly Correlated Fermionic-Systems Theoretical Quantum Simulation | Optimal 2D Fermionic QFT | Dynamical Systems: Fermionic-systems simulation for models that suffer from the sign problem such as the Fermi-Hubbard model, frustrated spin systems, the t-J model, or lattice gauge theories. | First-Principle Quantum Simulation | Proven | LSQ | No | No | No | No | | Circuit Model | 162 | https://arxiv.org/abs/1711.053 95... Quantum algorithms to simulate many-body... | https://doi.org/10.1103/PhysRevAppli... | 15/11/2017 |

| Quantum algorithm | Main subroutine | Fundamental mathematical problem | Applications | Heuristic/ Proven | NISQ/ LSQ | Oracular | Promise | Sampling | Partial dequantization | Dequantization link | Computational model | Google-Scholar citations | Link to arXiv paper | Link to peer-reviewed paper | Date |
|---|---|---|---|---|---|---|---|---|---|---|---|---|---|---|---|
| QCA for (3+1)D Q. Electrodynamics | Primitive | Dynamical Systems: Simulating Q. electrodynamics Hamiltonians, (3+1)D Q. electrodynamics Hamiltonian simulation | First-Principle Quantum Simulation: (3+1)D Q. electrodynamics | Proven | LSQ | No | No | No | No | | Quantum Cellular Automata | 4 | https://arxiv.org/abs/22... | https://quantum-journal... | 06/05/2022 |
| Aharonov–Jones–Landau Algo. | Hadamard Test | Dynamical Systems: Obtaining an approximation of the Jones polynomial of any braid. | First-Principle Quantum Simulation of topological quantum field theory | Proven | LSQ | No | No | No | No | | Circuit Model | 358 | https://arxiv.org/abs/qu... | https://link.springer... | 09/11/2005 |
| Q. Algo. for Feynman Loop Integrals | Grover | Dynamical Systems, Combinatorics: Bootstrapping of the causal representation of multiloop Feynman integrals in the loop-tree duality formalism from the identification of all internal configurations that fulfill causality among the 2^n potential solutions. Combinatorics: Identifying directed acyclic graphs | Classical Scientific Computing: High-energy physics, Perturbative QFT. Evaluation of quantum fluctuations at higher orders in the perturbative expansion by means of multiloop Feynman diagrams | Proven | LSQ | No | Yes, only for complex topology whose number of states fulfilling causal-compatible conditions is much smaller than the total combinations of ... | No | No | | Circuit Model | 20 | https://arxiv.org/abs/21.05/703 | https://doi.org/10.1007/JHE P05%282022%2... | 18/05/2021 |
| Q. Simulation of Effective Field Theories for Collider Physics | QFT | Dynamical Systems: Low-energy effective-field-theory dynamics (soft collinear effective theory) | First-Principle Quantum Simulation: Hadronic jet simulation | Proven | NISQ | No | No | No | No | | Circuit Model | 44 | https://arxiv.org/abs/21.03897 | https://doi.org/... | 09/02/2021 |
| Q. Simulation of Deuteron Binding Energy | VQE | Dynamical Systems: Pionless effective field theory dynamics (deuteron binding-energy computation) | First-Principle Quantum Simulation: Nuclear-physics simulation | Heuristic | NISQ | No | No | No | No | | Circuit Model | 394 | https://arxiv.org/pdf/18.03897 | https://doi.org/... | 11/01/2018 |
| Q. Algo. of Final State Shower | Primitive | Dynamical Systems: Simulating parton showers of high-energy scattering processes | First-Principle Quantum Simulation: High-energy physics (high-energy scattering amplitudes) | Proven | NISQ | No | No | Yes | No | | Circuit Model | 110 | https://arxiv.org/abs/19.04.03196 | | 05/04/2019 |
| Continuous-Variable QITE Formulation of Scalar Q. Field Theory | QITE, Q. Lanczos | Dynamical Systems: Computing Hamiltonian ground states and excited states | First-Principle Quantum Simulation: Scalar QFT simulation over 1D lattice | Heuristic | LSQ | No | No | No | No | | Linear Optical Quantum Computing | 22 | https://arxiv.org/abs/21... | https://doi.org/... | 02/07/2021 |
| QITE for Feynman-Kac Partial Differential Equation | QITE | Dynamical Systems: Solving Feynman-Kac equation | Classical Scientific Computing: Heat-physics simulation, quantum-system simulation. Operational Research: Finance (option pricing with stochastic volatility) | Heuristic | NISQ | No | No | No | No | | Circuit Model | 20 | https://arxiv.org/abs/21.2022-06... | https://doi.org/10... | 24/08/2021 |
| Variational Q. Algo. for Parton Distribution Functions | VQE | Dynamical Systems: Computing parton distribution functions in QCD | First-Principle Quantum Simulation: Quantum-chromodynamics computation | Heuristic | NISQ | No | No | No | No | | Circuit Model | 43 | https://arxiv.org/pdf/20.11.13934... | https://doi.org/10... | 27/11/2020 |
| Q. Algo. for Generating the Ground State of a Quantum Field Theory | Primitive | Dynamical Systems: Generating an approximation for the ground state of a quantum field theory | First-Principle Quantum Simulation: Simulating a massive scalar bosonic QFT on a quantum computer | Proven | LSQ | No | No | No | No | | Circuit Model | 10 | https://arxiv.org/abs/21.05708 | https://doi.org/10... | 12/10/2021 |
| Q. Many-Body Simulation on Fock Space | Primitive | Dynamical Systems: Mapping the many-body system onto an unconventional Anderson model on a complex Fock space network of many-... | First-Principle Quantum Simulation: Simulation of a many-body system in real space. | Proven | NISQ | No | No | No | No | | Circuit Model | 4 | https://arxiv.org/pdf/22.11.05603 | https://www.nature.com/artic... | 10/11/2022 |
| Q. Algo. for Matched Filtering | (Approximate) Q. Counting | Stochastic Processes & Statistics, Combinatorics: Comparing templates to the data and evaluate the matching | Classical Scientific Computing: Data analysis (ex: gravitational-wave detection). | Proven | LSQ | Yes | No | No | No | | Circuit Model | 9 | https://arxiv.org/abs/09.01535 | https://journals.aps.org/pres... | 03/09/2021 |
| Variational Q. Algo. for Poisson Equation | VQE | Dynamical Systems: Solving Poisson equation | Classical Scientific Computing: Variational quantum solver | Heuristic | NISQ | No | No | No | No | | Circuit Model | 60 | https://arxiv.org/pdf/17.09.10417... | https://journals.aps.org/pra/a bstract/10... | 13/12/2020 |
| Q. Simulation of Hydrodynamics via Madelung Transformation | QPE | Dynamical Systems: Solving hydrodynamical PDEs discretized as quantum lattice gas automata | Classical Scientific Computing: First-Principle Quantum Simulation: Solvings hydrodynamical PDEs | Proven | NISQ | No | No | No | Primitive is | | Circuit Model; Quantum Walks | 7 | https://arxiv.org/pdf/22.11... | | 11/05/2023 |
| Efficient Q. Algo. for Dissipative Nonlinear DEs | QLSA Childs et al. | Dynamical Systems: Solving dissipative non-linear equations | Classical Scientific Computing: Computational fluid dynamics, (middle-scale) reaction simulation, epidemiology, numerical modeling | Proven | LSQ | No | Yes, nonlinearity ratio R<1 | No | Primitive is | | Circuit Model | 126 | https://arxiv.org/abs/20.11.03185 | https://doi.org/10.1073/pna... | 06/11/2020 |

| Quantum algorithm | Main subroutine | Fundamental mathematical problem | Applications | Heuristic/ Proven | NISQ/ LSQ | Oracular | Promise | Sampling | Partial dequantization | Dequantization link | Computational model | Google-Scholar citations | Link to arXiv paper | Link to peer-reviewed paper | Date |
|---|---|---|---|---|---|---|---|---|---|---|---|---|---|---|---|
| Kacewicz' Q. Ordinary-Differential-Equation Algo. | QAE | Dynamical Systems: Solving ordinary differential equation in an optimal way | Classical Scientific Computing: Solving ODE | Proven | LSQ | Yes; driven function operator for f(t) | No | No | No | | Purely mathematical | 60 | https://arxiv.org/abs/quant-ph | https://www.sciencedirect.com | 06/10/2005 |
| Q. Algo. for Navier–Stokes Equation | Kacewicz' Q. Ordinary-Differential-Equation Algo. | Dynamical Systems: Solving Navier-Stokes equation | Classical Scientific Computing: Computational mechanics | Proven | LSQ | No | No | No | No | | Circuit Model, Quantum Walks | 31 (arXiv + peer-reviewed) | https://arxiv.org/abs/2104.04 | https://worldscientific | 05/03/2021 |
| Q. Hartree-Fock Variational Method | VQE | Dynamical Systems: Solving the time-independent Schrödinger equation | First-Principle Quantum Simulation: Quantum chemistry simulation using Hartree-Fock method. | Heuristic | NISQ | No | No | No | No | | Circuit Model | 527 | https://arxiv.org/abs/21 Hartree-Fock on a superconducting | https://doi.org/10.1126/scie | 08/04/2020 |
| Iterative QPE | Hadamard Test | Dynamical Systems: Finding the eigenvalue of a unitary operator | First-Principle Quantum Simulation: Quantum-chemistry simulation of particle spectrums | Proven | NISQ | No | No | No | No | | Circuit Model | 179 | https://arxiv.org/abs/qu ant-ph | https://doi.org/ | 25/10/2006 |
| Bayesian QPE | Primitive | Linear Algebra: Estimation of the eigenvalue of an eigenvector of a unitary operator. Dynamical Systems: Finding the spectrum of a Hamiltonian. | First-Principle Quantum Simulation | Heuristic | NISQ | No | No | No | No | | Circuit Model | 0 | 2303.0151 7.pdf (arxiv.org) | Not published | 02/03/2023 |
| Subspace-Search VQE | VQE | Dynamical Systems: Computes excited states of a given Hamiltonian | First-Principle Quantum Simulation: Computing excited states of a Hamiltonian | Heuristic | NISQ | No | No | No | No | | Circuit Model | 250 | 1810.0943 4.pdf (arxiv.org) | https://journals.aps. | 22/10/2018 |
| Variational Q. Algos. for Discovering Hamiltonian Spectra (Excited States) | QITE, Swap Test | Dynamical Systems: Computes excited states of a given Hamiltonian | First-Principle Quantum Simulation: Finding the excited states of molecular systems | Heuristic | NISQ | No | No | No | No | | Circuit Model | 220 | 1806.057 071 (arxiv.org) | https://journals.aps. Variational | 14/06/2018 |
| Variational Q. Algos. for (Many-Body) Green's Funtions and Spectrum Functions | Subspace-Search VQE | Dynamical Systems: Computes the Green's function of a given system | First-Principle Quantum Simulation: Computes the Green's funtion and spectrum function of a quantum system | Heuristic | NISQ | No | No | No | No | | Circuit Model | 79 | [1909.122 50] Calculatio n of the Green's | https://journals.aps. org/press | 26/09/2019 |
| Q. Algo. for Full Configuration Interaction (FCI) | Bayesian QPE | Dynamical Systems: Computes the ground-state energy of a many-body quantum eletronic system | First-Principle Quantum Simulation: Simulation of many-body fermions systems | Heuristic | NISQ | No | No | No | No | | Circuit Model | 11 | No Arxiv paper | https://pubs.acs. org/doi/abs | 08/11/2021 |
| Q. Algo. for Density Functional Theory (DFT) | HHL, VQE | Dynamical Systems: Density functional theory | First-Principle Quantum Simulation: Simulation of many-body fermions systems | Heuristic | LSQ | No | No | No | No | | Circuit Model | 2 | https://arxiv.org/abs/22 04.0444 | https://scipost.org/10. | 04/04/2022 |
| Adiabatic Q. Algo. for Artificial Graphene | QAOA | Dynamical Systems: Fermi-Hubbard-model simulation | First-Principle Quantum Simulation: Material-structure simulation | Proven | LSQ | No | No | No | No | | Adiabatic Quantum Computation, Circuit Model | 5 | https://arxiv.org/abs/22 04.030 13] | | 06/04/2022 |
| Variational Protein-Folding Simulation | VQE | Dynamical Systems: Simulating a coarse-grained model for protein folding | First-Principle Quantum Simulation: Simulation of protein folding | Heuristic | NISQ | No | Yes; only for branched heteropolymers comprised of N monomers on a tetrahedral (or "diamond") lattice | No | No | | Circuit Model | 72 | https://arxiv.org/abs/19 08.02163 | https://www.nature.com/artic les/s4153 4-021- 00368-4 | 06/08/2019 |
| Q. Algo. for Alchemical Optimization | VQE | Dynamical Systems: Finding the eigenspectrum of a quantum system's Hamiltonian | First-Principle Quantum Simulation: Design of molecular systems that best fit in a given external potential. Material design, drug design. | Heuristic | NISQ | No | No | No | No | | Circuit Model | 22 | https://arxiv.org/abs/20 08.06449 | https://pubs. rsc. org/en/co ntent/artic lehtml/20 | 16/10/2020 |
| Dirac-Equation QW | Primitive | Dynamical Systems: Dynamical evolution of the Dirac and Weyl equations | First-Principle Quantum Simulation: Quantum simulating the Dirac and the Weyl equations | Proven | LSQ | No | No | No | No | | Quantum Walks | 262 | https://arxiv.org/abs/he | https://journals.aps. | 19/04/1993 |
| QW for Non-Abelian Gauge Theory | Variant of Dirac-equation QW | Dynamical Systems: Dynamical evolution of a non-Abelian gauge theory | First-Principle Quantum Simulation: Quantum simulating non-Abelian gauge theories | Proven | LSQ | No | No | No | No | | Quantum Walks | 69 | https://arxiv.org/abs/16 | https://journals.aps. | 19/06/2016 |
| QW for Fermions in Curved Spacetime | Variant of Dirac-Equation QW | Dynamical Systems: Dynamical evolution of fermions in curved spacetime | First-Principle Quantum Simulation: Quantum simulating fermions in curved spacetime | Proven | LSQ | No | No | No | No | | Quantum Walks | 40 | https://arxiv.org/abs/16 | https://www.mdpi.com | 01/08/2017 |

| Quantum algorithm | Main subroutine | Fundamental mathematical problem | Applications | Heuristic/ Proven | NISQ/ LSQ | Oracular | Promise | Sampling | Partial dequantization | Dequantization link | Computational model | Google-Scholar citations | Link to arXiv paper | Link to peer-reviewed paper | Date |
|---|---|---|---|---|---|---|---|---|---|---|---|---|---|---|---|
| QWs Meet LGT | Variant of Dirac-Equation QW | Dynamical Systems: Unitary dynamical evolution of fermions on lattices | First-Principle Quantum Simulation: Quantum simulating fermions on lattices | Proven | LSQ | No | No | No | No | No | Quantum Walks | 1 | https://arxiv.org/abs/2208.14907 | https://iopscience.iop.org/article | 28/12/2022 |


 % \caption{Classification table}
  %\label{fig:Classification table}
\end{figure}

\end{document}